\documentclass[aps,prf,onecolumn,citesort]{revtex4-2}
\usepackage{graphicx}
\usepackage{epstopdf, epsfig}
\usepackage{amsmath}
\usepackage{amssymb}
\usepackage{xcolor}
\usepackage{color}
\usepackage{bm}
\usepackage[bookmarks,colorlinks]{hyperref}
\pdfoutput=1

\newcommand{\ec}[1]{\textcolor{black}{#1}}

\begin{document}
\title{Convective heat transfer in the Burgers-Rayleigh-B\'enard  system}
\author{Enrico Calzavarini}
\email{enrico.calzavarini@univ-lille.fr}  
\affiliation{Universit$\acute{e}$ de Lille, Unit$\acute{e}$ de M$\acute{e}$canique de Lille - J. Boussinesq, UML ULR 7512, F 59000 Lille, France}
\author{Silvia C. Hirata}
\affiliation{Universit$\acute{e}$ de Lille, Unit$\acute{e}$ de M$\acute{e}$canique de Lille - J. Boussinesq, UML ULR 7512, F 59000 Lille, France}


\date{\today}


\begin{abstract}
The dynamics of 
heat transfer in a model system of Rayleigh-B\'enard (RB) convection reduced to its essential, here dubbed Burgers-Rayleigh-B\'enard (BRB), is studied. The system is spatially one-dimensional, the flow field is compressible and its evolution is described by the Burgers equation forced by an active temperature field. The BRB dynamics shares some remarkable similarities with realistic RB thermal convection in higher spatial dimensions: i) it has a supercritical pitchfork instability for the onset of convection which solely depends on the Rayleigh number $(Ra)$ and not on Prandlt $(Pr)$, occurring at the critical value $Ra_c = (2\pi)^4$ ii) the convective regime is spatially organized in distinct boundary-layers and bulk regions, iii) the asymptotic high $Ra$ limit displays the Nusselt and Reynolds numbers scaling regime $Nu = \sqrt{RaPr}/4$ for $Pr\ll 1$, $Nu=\sqrt{Ra}/(4\sqrt{\pi})$ for $Pr\gg1$ and
$Re = \sqrt{Ra/Pr}/\sqrt{12}$, thus making BRB the simplest wall-bounded convective system exhibiting the so called ultimate regime of convection. These scaling laws, derived analytically through a matched asymptotic analysis are fully supported by the results of the accompanying numerical simulations. 
A major difference with realistic natural convection is the absence of turbulence. The BRB dynamics is stationary at any $Ra$ number above the onset of convection. This feature results from a nonlinear saturation mechanism whose existence is grasped by means of a two-mode truncated equation system and via a stability analysis of the convective regime.
\end{abstract}
\maketitle

\section{Introduction}
Thought experiments, toy-models, low-dimensional representations are keys to the scientific thinking, and allow to get insight into the complex physics of many real systems.  In fluid-dynamics research,  reduced models obtained, e.g., via expansion and truncations of the original dynamical equations have been used to conceptualize and to understand, for instance, the chaotic dynamics of flows (Lorenz system \cite{Lorenz1963}), or the physics of energy cascade in developed turbulence (Shell models \cite{Gledzer1973,OhkitaniYamada1989}). In this study we focus on the problem of thermal convection and, in the spirit of the one-dimensional (1D) toy model for granular media introduced by Du, Li and Kadanoff \cite{DuLiKadanoff1995, Eshuis2009}, we here introduce a stripped-down mock-up of the classical Rayleigh-B\'enard (RB) system \cite{Rayleigh1916}. 

The RB has been extensively studied either in its spatial three-dimensional or in its two-dimensional version \cite{Getling,RevModPhys.81.503}. We are not aware of any study of the system in one-dimension. This is after all quite understandable, since in 1D the incompressibility condition for the flow does not hold and one expects a rather different physical behaviour.  This is indeed already the case for the 1D version of the Navier-Stokes equation, i.e., the Burgers equation \cite{Burgers1948}. It is well known that the Burgers equation does not display a turbulent behaviour, because it can be recast in term of a diffusion equation via the Hopf-Cole transformation \cite{Hopf1950,Cole1951}.
However, stochastically forced Burgers equation does produce a special kind of turbulence, dubbed Burgulence \cite{Frisch2001}, that has drawn the attention of recent research \cite{BEC20071}.\\
We show in this study that a 1D deterministically-forced version of the RB system that we dub Burgers-Rayleigh-B\'enard (BRB) system can be defined. Interestingly, this system possesses a certain number of similarities with thermal convection in higher spatial dimensions:  it has a supercritical linear instability for the onset of convection, the convective regime is spatially organized in distinct boundary-layers and bulk regions, and the asymptotic high $Ra$ limit displays the so-called ultimate Nusselt and Reynolds numbers scalings \cite{RevModPhys.81.503}, although it lacks of any turbulent behaviour.
It also admits shock-like solutions that are peculiar of the Burgers dynamics.    

The article is organized as follows. 
We first define the BRB system, next we examine its most relevant symmetries and its global properties. In particular, we introduce the definition of the Nusselt and Reynolds numbers, which are the two global response parameters of the system. 
Secondly, we perform a theoretical analysis on the system dynamics, focusing on the calculation of the linear instability threshold for convection, on the subsequent non-linear saturation mechanism and on a derivation of a steady matched asymptotic solution for the very intense convection state. Third, we push forward the analysis by means of a numerical approach. In particular we show that the system is stationary at all Rayleigh numbers, this is first revealed empirically then verified by means of numerically-based stability analysis. We then show that the Nusselt and Reynolds number asymptotically approaches the ultimate state of thermal convection, this both in their Rayleigh and Prandtl number dependencies.  
Finally, we discuss the implications of our findings and possible perspectives.

\section{The Burgers-Rayleigh-B\'enard model system}
\subsection{Equations of motion}
We study the spatio-temporal evolution of a single-component velocity $W(Z,\tau)$ and temperature $T(Z,\tau)$ fields in a one-dimensional domain $Z\in \left[0,H \right]$, described by the coupled system of differential equations:
\begin{eqnarray}
 W_{\tau} + W\ W_Z &=& \nu \ W_{ZZ} +   \beta g (T -T_c) \label{eq:burgersT}\\
T_{\tau} + W\ T_Z &=& \kappa \ T_{ZZ} \label{eq:T},
\end{eqnarray}
with Dirichlet boundary conditions 
\begin{eqnarray}
W &=& 0 ,\ T = \tfrac{\Delta}{2}\quad  \textrm{in} \ Z=0\ (\textrm{bottom}),\\ 
W &=& 0,\ T = -\tfrac{\Delta}{2} \quad  \textrm{in} \  Z=H\ (\textrm{top}),
\end{eqnarray}
where $\nu$ and $\kappa$ denotes respectively the viscosity and thermal diffusivity,  $\beta$ the  thermal expansion coefficient, $g$ the gravitational acceleration intensity, and $T_c$ the linear profile given by $T_c(Z) = -(\Delta/H)Z+\Delta/2$, which is also said conductive because it represents a solution for the temperature field when $W=0$ in all the domain. Furthermore, to keep the similarity with realistic RB convection we adopt the additional constraint that the global value of velocity and temperature fields are null, 
\begin{equation}
\int_0^H W\ dZ = \int_0^H T\ dZ = 0  \  \  (\textrm{no-zero mode condition}).
\end{equation}
This prevents the possibility for the system to acquire a vertical mean flow and to heat up/cool off. We will comment later on the consequence of this constraint.\\
As  we have already mentioned, the above model constitutes an \ec{oversimplified} representation of the Rayleigh-B\'enard  system. It can be loosely obtained from the Navier-Stokes-Boussinesq set of equations for the three-dimensional velocity $\textbf{U} = (U,V,W)$ and temperature  $T$, by assuming that (i) the vertical component of the velocity, $W$, and the temperature depend only on the vertical direction, $Z$, (ii) by removing the hydrodynamic pressure field and (iii) by expressing the buoyancy force as proportional to the temperature deviation from the local conductive temperature profile.
With the above assumptions, the equations for $T$ and $W$ decouple from the ones for horizontal components $U,V$ and can be treated separately.  As a consequence the vertical velocity gradient $W_Z$ becomes unconstrained and the corresponding unidimensional velocity field is compressible.
\ec{We stress that the BRB model can not be regarded as a low-dimensional mean-field form of the Boussinesq system, neither a model of convection in compressible gases (where the continuity equation would have a different form). However, we believe that despite its incompleteness this model  is useful to get an insight in what does/does not occur in the realistic system.}

The equations (\ref{eq:burgersT}-\ref{eq:T}) can be made dimensionless by means of the linear size of the domain (or height $H$), the free-fall velocity $U_f=\sqrt{\beta g H \Delta}$ and the global  temperature gap $\Delta$ (i.e. the difference between the top temperature and the bottom one). This leads to the two control parameters in the system: the Rayleigh number $Ra = (U_f H)^2/(\nu \kappa)$ and the Prandtl number $Pr = \nu/\kappa$. 
With these choices the equations can be conveniently rewritten in term of the velocity, $w =W/U_f$, and the temperature deviation from the conductive profile, $\theta(z,t)  = (T(Z,\tau)-T_c(Z))/\Delta$, as:
\begin{eqnarray}
 w_t + w\ w_z &=& \sqrt{\tfrac{Pr}{Ra}}\ w_{zz} +   \theta \label{eq:burgers}\\
\theta_t + w\ \theta_z &=& \tfrac{1}{\sqrt{PrRa}}\ \theta_{zz} +  w \label{eq:theta},
\end{eqnarray}
with 
$w = \theta = 0\quad  \textrm{in} \quad z=0\  \textrm{and}\  z=1$ and $\langle w \rangle = \langle \theta \rangle = 0$, where $\langle \ldots \rangle= \int_{0}^{1}\ldots dz$ is the spatial average (all lower-case letter denote dimensionless variables).
Equation (\ref{eq:burgers}) is the 1D forced Burgers equation, which is coupled to the advection-diffusion equation for a scalar field   (\ref{eq:theta}) which is in turn forced by $w$. 
\begin{figure*}[htb]
	\begin{center}
		\includegraphics[width=0.85\textwidth]{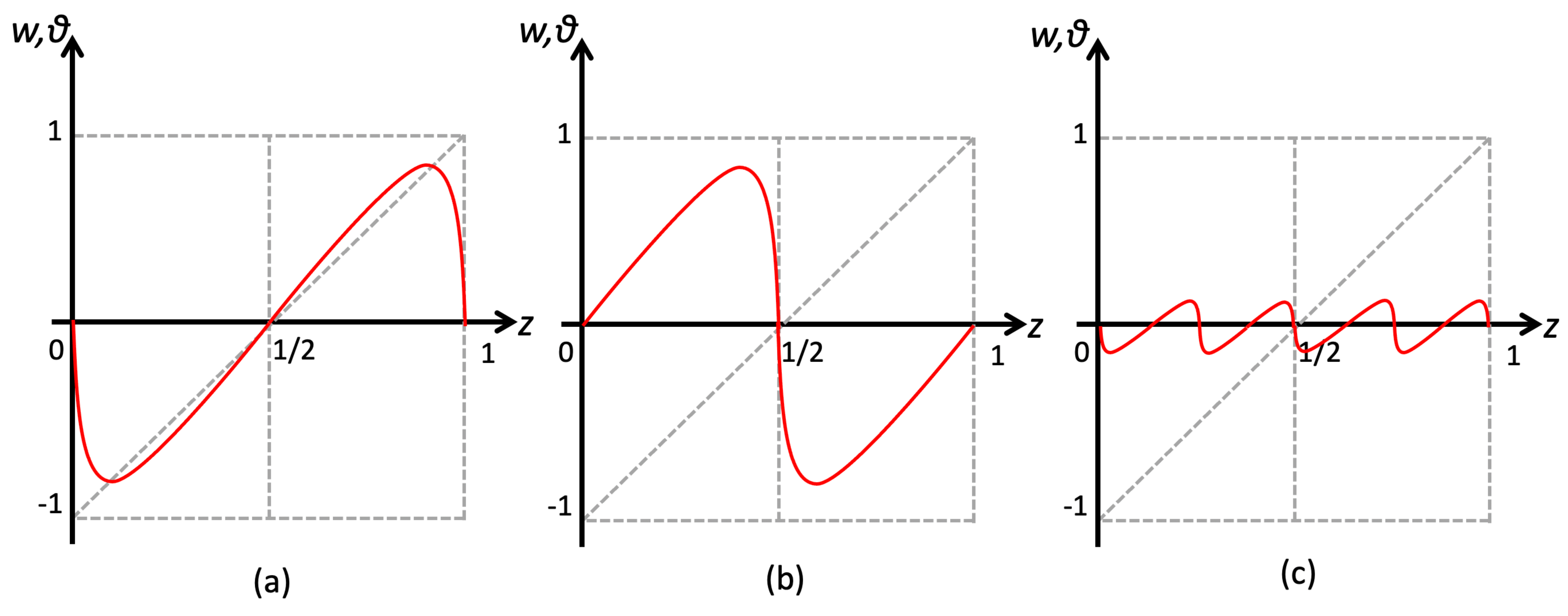}
		\vspace{-0.5cm}
		\caption{Symmetries of the system of equations: (a) odd symmetry of $w$ and $\theta$ with respect to the position $z=1/2$; (b) \ec{swap symmetry with resect to the system mid-point (\ref{eq:thirdsym}); (c) rescaling transformation (\ref{eq:rescaling1})-(\ref{eq:rescaling2}) with $n=1$}.}
		\label{fig:sym}			
	\end{center}
\end{figure*}	

\subsection{Symmetries}\label{sec:sym}
The system (\ref{eq:burgers}-\ref{eq:theta}) enjoys a series of symmetries which greatly affect its dynamics. We describe them in detail in this section. To begin with we note that when $Pr=1$, $\theta=w$ is a permitted solution of the BRB model system.

Second, the set of equations (\ref{eq:burgers}-\ref{eq:theta})  is invariant with respect to the transformation $(z,\theta) \to (1-z, -\theta)$ which also implies $w\to -w$ because by definition $w=dz/dt$.
This means that if the couple $\theta(t,z), w(t,z)$ indicates a solution of the equations, than also 
$- \theta(t,1-z), - w(t,1-z)$ is a solution. Combining this with the condition $\langle w \rangle = \langle \theta \rangle = 0$, it entails that both $\theta$ and $w$ are odd functions with respect to $z=1/2$, and so $\theta(z=1/2) = w(z=1/2)=0$.

A third symmetry is the following:
\begin{equation}\label{eq:thirdsym}
z \to 
\begin{cases}
 z+1/2 \ \textrm{if}\ z \leq1/2\\
 z-1/2 \ \textrm{if}\ z>1/2
 \end{cases}
 \  \  \textrm{or} \quad 
 z \to z+\textrm{sign}\left(\frac{1}{2}-z \right) \frac{1}{2}.
 \end{equation}
It corresponds to swapping the spatial interval $[0,1/2]$ with the one $[1/2,1]$. As shown in the sketch in Fig.~\ref{fig:sym}(a)-(b) this symmetry transforms what we call a ``boundary-layer" type solution to a ``shock" type solution (more on this later). In other words, due to the zero boundary conditions and to the second symmetry, the functions $w(z)$ and $\theta(z)$ can be seen as periodic odd functions. This means that adding a phase of half the period is still a solution of the system. Equivalently one can say that, $z\to z+1/2$ is a symmetry of the system.

A fourth remarkable symmetry of the system is the following: Let's call $w(t,z;Ra,Pr) , \theta(t,z;Ra,Pr)$ the solution of the system for a given value of the parameters $Ra,Pr$. The system is then invariant with respect to the transformation:
\begin{eqnarray}\label{eq:rescaling1}
w(t,z;Ra,Pr)  &\to&  w(t,2 n z;\tfrac{Ra}{2n},Pr)/(2n)\\
\label{eq:rescaling2}
\theta(t,z;Ra,Pr) &\to& \theta(t,2 n z;\tfrac{Ra}{2n},Pr)/(2n)
\end{eqnarray}
where $n$ is a positive integer number.  This ``rescaling transformation" symmetry is illustrated in Fig. \ref{fig:sym}(c) \ec{for the case $n=1$}.\\

\subsection{Global response parameters: Nusselt and Reynolds}

To derive the expression for the global heat flux it is convenient to resort to the dimensional notation. 
The temperature equation (\ref{eq:T}) in conservative form reads: $T_{\tau} + (J_{T})_Z = 0$, where
\begin{equation}\label{eq:flux}
J_{T}(Z,\tau) =  W T - \kappa T_{Z}  -  \int_{0}^Z  T W_{Z'}\ dZ',   
\end{equation}
is the local and instantaneous heat flux at position $Z$ and time $\tau$. Averaging the conservative form equation over time (here denoted as overline) and assuming a steady state gives the expression of the mean global heat-flux:
\begin{equation}\label{eq:mean-flux}
\overline{J_{T}}(Z)=   \overline{W T} - \kappa  \overline{T}_{Z}  -  \int_{0}^Z  \ \overline{T W_{Z'}}\ dZ'   = const.
\end{equation}
The integral term in the above expression, which is absent in the mean heat flux of the RB system, is a consequence of the compressibility of the velocity field.
The mean Nusselt number is defined by adimensionalizing the mean global heat flux with respect to the conductive heat flux  (i.e. the state where $W=0$ and $T=T_c$): 
\begin{eqnarray}
 Nu &\equiv& \frac{\overline{J_{T}}  }{J_{T_c}}  = const.\label{eq:nu_def0}
\end{eqnarray}
We observe that by plugging into the above expression the dimensionless temperature fluctuation $\theta$, and evaluating the expression either in $z=0$ or $z=1$, gives the following equivalent expressions for $Nu$:
\begin{eqnarray}
 Nu &=& 1 - \overline{\theta}_{z}(0) = 1 - \overline{\theta}_{z}(1). 
\label{eq:nu_def1}
\end{eqnarray}
One can remark that this same expression for $Nu$ is obtained in the RB flow ruled by the Boussinesq system of equations. On the contrary, if one considers the spatial average of $Nu$ (spatial average of eq. (\ref{eq:nu_def0})) one gets:
\begin{eqnarray}
 Nu = 1 + \sqrt{PrRa}\left( \langle \overline{w \theta} \rangle -  \langle \int_{0}^z  (\overline{\theta w_{z'}} + \overline{w}) dz' \rangle \right)
\label{eq:nu_def2}
\end{eqnarray}
which is different from the RB expression by the appearance of the integral term on the \textit{r.h.s.}, which originates, as already mentioned, by the flow compressibility.
The volume averaged expression of the Nusselt number is convenient for numerical calculations, as it is less affected by discretization and numerical errors (we will use this expression in the numerical calculations presented in this article).\\ 
Finally, we note that the Reynolds number defined as a system response parameter is here:
\begin{equation}
Re \equiv \frac{\overline{\langle W^2 \rangle}^{1/2} H }{\nu} = \sqrt{\frac{Ra}{Pr}}\ \overline{\langle w^2 \rangle}^{1/2}. 
\end{equation}

\section{The BRB dynamics: theoretical analysis}
This section presents some notable analytical results on the dynamics of the BRB model system. First, we perform the linear stability analysis to determine the transition from the conductive to the convective state. Second, we address the non-linear saturation mechanism that is responsible for the stabilization of the flow after the inception of convection. Third, by means of a standard matched asymptotic ($ma$) analysis, we solve the BRB system of equations in steady condition in the limit of large $Ra$ numbers. Finally, based on the $ma$ solution we derive the asymptotic-in-Rayleigh scaling laws for the Nusselt and Reynolds numbers.

\subsection{Onset of convection}\label{sec:stability}
The linearization  of the system (\ref{eq:burgers}-\ref{eq:theta}) with respect to $w$ and $\theta$ satisfies solutions of the form 
\begin{equation}
w = \Sigma_{n=1}^{\infty} w_n e^{\sigma_n t} \sin(n k z), \quad
\theta = \Sigma_{n=1}^{\infty} \theta_n e^{\sigma_n t} \sin(n k z),
\end{equation}
where $k=2\pi$ and $n$ is an integer value, and with the growth rate
\begin{equation}
\sigma_n =  \frac{\sqrt{(Pr+1)^2(n k)^4+4Pr(Ra-(n k)^4)} -(Pr+1)(n k)^2 }{2 \sqrt{Ra Pr}}.
\label{eq:sigma}
\end{equation}
Therefore,  for $Ra > Ra_c = k^4 = (2\pi)^4 \simeq 1558$ and at any $Pr$ value the system becomes linearly unstable ($\sigma_1>0$).
The critical Rayleigh number happens to be the same as in the three-dimensional three-periodic homogeneous Rayleigh-B\'enard system \cite{CalzavariniPF2005,CalzavariniDoeringPRE2006} although in that case the perturbation form is different as it depends only on the horizontal coordinates.
We observe that the relative amplitude of the velocity and temperature field is $w_1/ \theta_1 = (\sigma_1 + k^2/\sqrt{Pr Ra})$, this implies that   $w_1 = \theta_1$ for $Pr=1$.
This prediction will be verified in Sec. \ref{sec:numerics} by means of a numerical simulation
 starting from a tiny white noise perturbation on $w$ and $\theta$ fields (see also Fig.\ref{fig:nu_ra}).
 
The described exponentially growing solution is eventually saturated by the presence of the nonlinear terms, as we discuss in the next section. 

\begin{figure}[htb]
\begin{center}
		\includegraphics[width=0.55\textwidth]{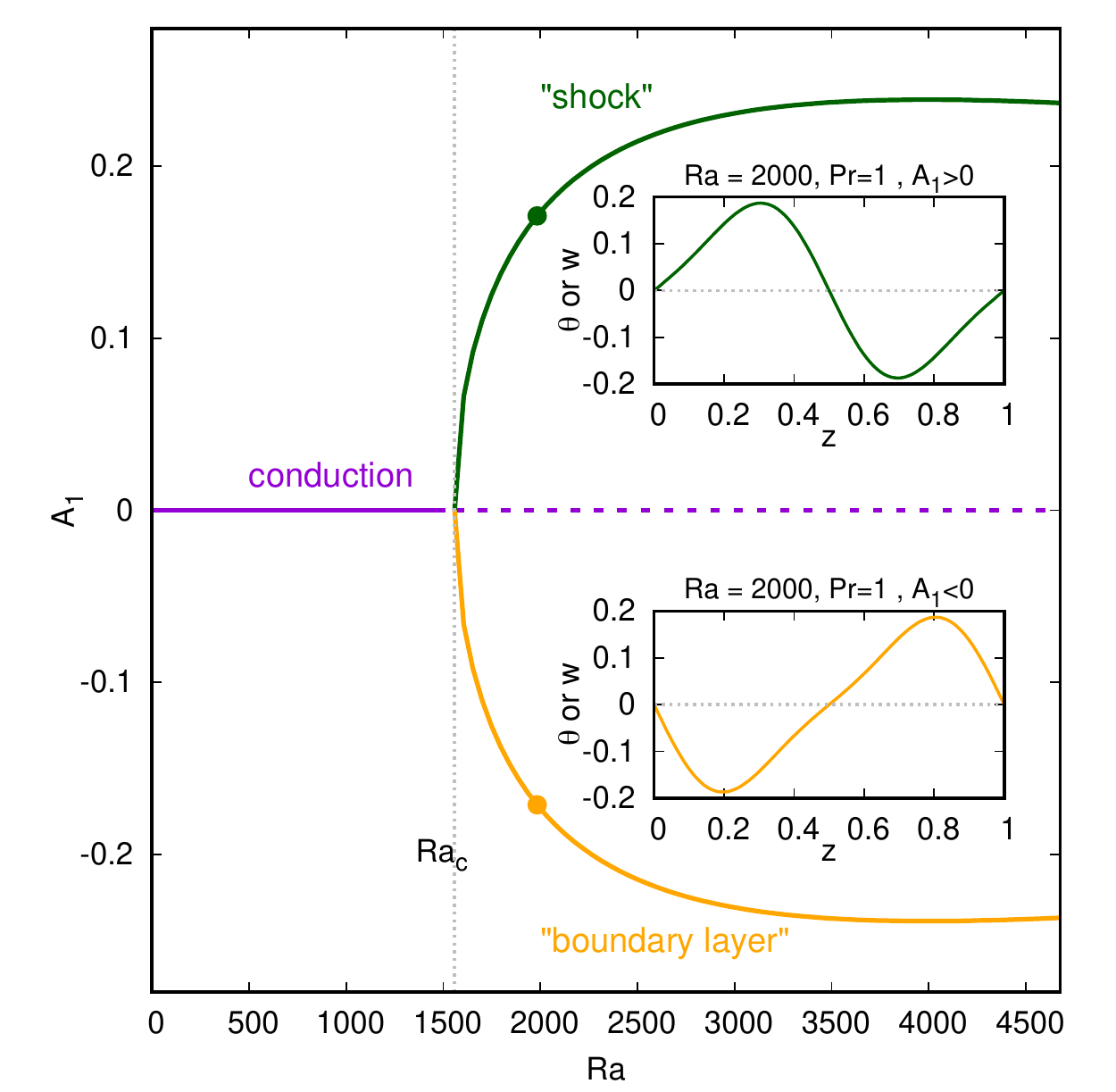}
		\vspace{-0.25cm}
		\caption{Steady solution of the amplitude $A_1$ of the two-mode truncated system as a function of the control parameter $Ra$, as given in eq. (\ref{A1cond}),(\ref{A1conv}). The graph has a typical structure of a supercritical pitchfork bifurcation, solid lines denote stable branches while the dotted one denotes the unstable brach. We indicate as ``conduction" the solution corresponding to $A_1=0$ (and $A_2=0$ too), with ``shock" type solution the one presenting a sharp variation on the bulk of the system, which corresponds to the case $A_1>0$, and finally with ``boundary layer" type the one where the variation occurs close to the boundaries ($A_1<0$ in this case). The insets show the complete steady two-mode solutions for $\theta(z)$ or $w(z)$ of the shock and boundary layer types for the case $Ra=2000$, marked with colored circles on the main panels.}
		\label{fig:fork}			
	\end{center}
\end{figure}	

\subsection{Non-linear saturation mechanism}

Similarly to what occurs in a three-dimensional RB system in slightly supercritical conditions ($Ra \gtrsim Ra_c$) the exponential growth rate of the perturbation rapidly saturates into a convective steady state. This phenomenology can be promptly explained  for the BRB at  $Pr=1$ by means of a two modes Galerkin expansion, which we detail in the following.
We assume 
\begin{equation}\label{eq:galerkin}
w(z,t) = \Sigma_{n=1,2} A_n(t) \sin{(n k z)}, \quad 
\theta(z,t) = \Sigma_{n=1,2} B_n(t) \sin{(n k z)}.
\end{equation}
Upon its substitution into the equations of motion, retaining only terms in $\sin{(k z)}$ and $\sin{(2 k z)}$, we obtain the first-order differential system for the evolution of the amplitudes of the four considered modes:
\begin{eqnarray}
\!\!\!\!\!\!\dot{A_1}  &=&  \ \frac{Ra_c^{1/4}}{2}A_1 A_2     - \sqrt{\frac{Ra_c}{Ra/Pr}}   A_1 + B_1\\
\!\!\!\!\!\!\dot{A_2} &=&   -\frac{Ra_c^{1/4}}{2} A_1^2          -  4 \sqrt{\frac{Ra_c}{Ra/Pr}}  A_2 + B_2\\
\!\!\!\!\!\!\dot{B_1}  &=&  \ \frac{Ra_c^{1/4}}{2}(2A_1 B_2  -  A_2 B_1)   - \sqrt{\frac{Ra_c}{Ra Pr}}  B_1 +\! A_1\\
\!\!\!\!\!\!\dot{B_2} &=&   -\frac{Ra_c^{1/4}}{2} A_1 B_1      -  4 \sqrt{\frac{Ra_c}{Ra Pr}}  B_2 + A_2,
\end{eqnarray}
where we have taken into account that $Ra_c = k^4$. Such system can not be solved analytically, however that is possible for the special case $Pr=1$. In this condition, the equations for $A_i$ and $B_i$ become identical, hence the differential system reduces to a two-dimensional one:
\begin{eqnarray}
\label{eq:2modes1}
\dot{A}_1&=&\sigma_1 A_1+\frac{Ra_c^{1/4}}{2}A_1 A_2\\
\dot{A}_2&=&\sigma_2 A_2-\frac{Ra_c^{1/4}}{2}A_1^2, 
\label{eq:2modes2}
\end{eqnarray}
with $\sigma_1=1-\sqrt{Ra_c/Ra}>0$ and $\sigma_2=1-4 \sqrt{Ra_c/Ra}<0$ for $Ra_c<Ra<16 Ra_c$. As the characteristic time-scale of the second mode $\tau_2=|\sigma_2|^{-1}$ is smaller than the characteristic time of the first mode $\tau_1=|\sigma_1|^{-1}$, we can perform a so-called adiabatic elimination and take $\dot{A}_2\approx0$ in the vicinity of the bifurcation threshold. By doing so, we obtain the following Landau equation \cite{Landau1987Fluid} for the evolution of the amplitude $A_1$:
\begin{equation}
\dot{A}_1= \sigma_1 A1- \gamma A_1^3
\label{landau}
\end{equation}
with $\gamma=\frac{\sqrt{Ra_c}}{4(4\sqrt{\frac{Ra_c}{Ra}}-1)}$. Eq.~(\ref{landau}) admits three steady solutions: 
\begin{eqnarray}
A^{ss}_1 &=& 0 \mathrm{\:\:(conductive\: state)}
\label{A1cond}\\
A^{ss}_1 &=& \pm \sqrt{\sigma_1/\gamma} \mathrm{\:\:(two\ convective\:states)}.
\label{A1conv}
\end{eqnarray}
 Assuming the perturbation expansion $A_1(t)=A_1^{ss}+\varepsilon A_{1p}(t)+O(\varepsilon^2)$, the amplitude equation at order $O(\varepsilon)$ writes
\begin{equation}
\dot{A}_{1p}= A_{1p}(t) (\sigma_1-3 \gamma (A_1^{ss})^2)
\end{equation}
A solution of the linear and homogeneous equation above can be written as  $A_{1p}(t)=A_{1p}(0)\exp{(-i \omega t)}$, which leads to
\begin{equation}
\sigma_1-3 \gamma (A_1^{ss})^2+ i \omega=0
\end{equation} 
Since all the coefficients are real and $\omega=\omega_R+i \omega_I$, we get
$ \omega_R=0 $ (meaning that solutions are  stationary) and $\omega_I=\sigma_1-3 \gamma (A_1^{ss})^2$. 
Hence, the growth rates of the trivial (\ref{A1cond}) and non-trivial (\ref{A1conv}) steady-states are $\omega_I=\sigma_1$ and $\omega_I=-2 \sigma_1$ respectively. Therefore, for $Ra<Ra_c$ the trivial conductive solution is stable, whereas for $Ra>Ra_c$ the non-trivial convective solutions are the ones to be stable. 
This corresponds to a supercritical pitchfork bifurcation, as illustrated in the graph of  Fig. \ref{fig:fork}.\\
The multiplicity of the above convective solutions is a consequence of the swap symmetry mentioned in the previous section, which in the case of a sinusoidal profile takes the form $\sin(n k z) \to (-1)^n \sin(n k z)$ for integers $n$. In the two mode expansion it therefore affects only the mode $A_1$. This produces a change from a solution with sharp variations at the boundaries, which we can denote as ``boundary layer" type solution, to one with a sharp transition in the bulk of the domain, that we call ``shock" type solution. While the boundary layer type solution is analogous to the vertical flow profile observed in the RB system, the shock type solution has no corresponding in 2D or 3D convective systems. For this reason, in our analysis we will mainly focus on the former kind of solution.
The temperature/velocity spatial profiles corresponding to the complete convective steady solutions of (\ref{eq:2modes1})-(\ref{eq:2modes2}), i.e., the two-mode truncated series (\ref{eq:galerkin}) with  
\begin{eqnarray}\label{eq:galerkin-coeff1}
A_1 &=& \pm \sqrt{\sigma_1/\gamma} = \frac{\pm 2}{Ra_c^{1/4}}  \left(1  - \sqrt{\frac{Ra_c}{Ra}}  \right)^{\frac{1}{2}}\left( 4 \sqrt{\frac{Ra_c}{Ra}} - 1 \right)^{\frac{1}{2}}\\
\label{eq:galerkin-coeff2}
A_2 &=&  -2 \sigma_1/Ra_c^{1/4} = \frac{-2}{Ra_c^{1/4}} \left(1  - \sqrt{\frac{Ra_c}{Ra}}  \right)
\end{eqnarray}
are traced in Fig.\ref{fig:fork} (insets) \footnote{
The stability of the three fixed points can be also assessed by means of the analysis of the eigenvalues of the Jacobian of the system of equations (\ref{eq:2modes1}),(\ref{eq:2modes2}), and in particular their dependence with respect to the control parameter $Ra$. This leads to the fact that the equilibrium point $A_1 =A_2=0$ changes from stable to unstable when $Ra>Ra_c$, while the two fixed points (\ref{eq:galerkin-coeff1}),(\ref{eq:galerkin-coeff2}) become stable as long as $Ra_c < Ra < 16 Ra_c$. This identifies a supercritical pitchfork bifurcation, see Fig.\ref{fig:fork}}.\\
The above convective solutions allows to estimate the Nusselt number near to the onset, as
\begin{equation}\label{eq:galerkin-Nu}
Nu = 1 - 2\pi\left( A_1 + 2 A_2 \right).
\end{equation}
They also imply that 
\begin{equation}
\langle w^2 \rangle  =  \langle \theta^2 \rangle  =\langle w \theta \rangle = \frac{1}{2}( A_1^2 + A_2^2) = \frac{6}{\sqrt{Ra}}\left(1  - \sqrt{\frac{Ra_c}{Ra}}  \right),
\end{equation}
so the Reynolds number becomes:
\begin{eqnarray}\label{eq:galerkin-re}
Re &=& \sqrt{ 6 \left(\sqrt{Ra}  - \sqrt{Ra_c} \right).}
\end{eqnarray}
These predictions will be tested in Sec. \ref{sec:numerics} by means of direct numerical simulations.\\

Finally, we wish to comment on the role of the no-zero mode condition, $\langle w \rangle = \langle \theta \rangle = 0$, introduced in our model system. The removal of this condition allows for an anticipated onset of convection at $Ra_c = \pi^4$ (16 times smaller than in the present case). It also permits non-odd solutions characterized by a single boundary layer on one of the sides of the domain. By virtue of the system symmetries, these solutions are trivially linked to the ones described above: their period and their amplitude are doubled but all scaling properties remains identical.

\subsection{Large-Ra asymptotic dynamics}

What happens to the system dynamics at large $Ra$? Does it keep stationary or rather does it become time-dependent, and possibly chaotic and turbulent? Replying to these questions on the basis of a theoretical analysis is challenging. We will try to reply to this question with the aid of numerical simulations, in Sec \ref{sec:numerics}.
However, it is possible to derive a steady solution of (\ref{eq:burgers})-(\ref{eq:theta}) in the limit of large $Ra$ and at any $Pr$, by means of the standard technique of matched asymptotic expansion. In the following we detail the derivation of such remarkable solution and we will then use it to give a prediction for the Nusselt and Reynolds scaling in the asymptotically large $Ra$ regime.  The numerical simulations, described in Sec \ref{sec:numerics}, will prove that the steady $ma$ solution describe strikingly well the real behaviour of the BRB system.

\subsubsection{Approximate stationary solution with matched asymptotic expansion}
\ec{The use of matched asymptotics to describe the structure of shocks in the Burgers equation at very small viscosities is a classical approach \cite{BenderOrszag1999}. This technique has been applied in several studies to estimate the contribution of shocks to the anomalous dissipation of kinetic energy \cite{Goodman1992}, to the energy spectrum of solutions \cite{Boyd1991}, to propose closures for statistical theories of Burgers' turbulence \cite{Weinan1999} and more recently to explain the spontaneous stochasticity of Lagrangian trajectories in Burgers equation \cite{Eyink2015}.}\\

Here we look for a solution of Eq. (\ref{eq:burgers}-\ref{eq:theta}) under the assumption that
both $\tfrac{\sqrt{Pr}}{\sqrt{Ra}}$ and $\tfrac{1}{\sqrt{Ra Pr}}$ are small parameters (or equivalently $Ra^{-1} \ll Pr \ll Ra$). First, we consider the solution of the system far from boundaries, denoted as \textit{outer} solution. In this region the dissipative terms are negligible, because they multiply the above mentioned small parameters, hence the system reduce to:
\begin{eqnarray}
w\ w_z &=&    \theta \label{eq:stat_burgers_out}\\
w\ \theta_z &=&  w \label{eq:stat_theta_out},
\end{eqnarray}
From the second equation we get $\theta_{out} = z + c_1$, which is then plugged into the first equation leading to $w w_z = z+c_1$ which admits the solution $w_{out} = \sqrt{z^2 + 2 c_1 z + c_2}$. By using the condition that the solution should be an odd function in the domain (i.e. $w(1/2)=\theta(1/2)=0$) we determine the constants $c_1=-1/2$, $c_2=1/4$ and so $w_{out} = \theta_{out} = z - 1/2$.\\
Second, we consider the solution near to a boundary, denoted as \textit{inner} solution (we choose here the boundary close to $z=0$). In this region, the application of standard least degeneracy principle \cite{vandyke} leads to the system:
\begin{eqnarray}
w\ w_z &=&     \tfrac{\sqrt{Pr}}{\sqrt{Ra}}\ w_{zz}\label{eq:stat_burgers_in}\\
w\ \theta_z &=&   \tfrac{1}{\sqrt{PrRa}}\ \theta_{zz} \label{eq:stat_theta_in},
\end{eqnarray}
Here one first solve the equation for $w$. This leads to $w_{in}=-a\tanh{\left(a\sqrt{\frac{Ra}{Pr}}\ z / 2\right)}$ where we have adopted the boundary condition in  $w(0)=0$. The constant $a$ can be determined by matching the inner and outer solutions for $w$, $\lim_{z\to\infty} w_{in} =  \lim_{z\to 0} w_{out}$: $-a=-1/2$, so $a=1/2$.
\begin{equation}
w_{in}=-\frac{1}{2}\tanh{\left(\sqrt{\frac{Ra}{Pr}}\ \frac{z}{4} \right)}
\end{equation}
By substituting now the inner solution $w_{in}$ in the equation for $\theta$ we obtain 
\begin{equation}
-\frac{1}{2} \tanh{\left(\sqrt{\frac{Ra}{Pr}}\ \frac{z}{4} \right)} \ \theta_z =   \tfrac{1}{\sqrt{PrRa}}\ \theta_{zz} 
\end{equation}
which can be rewritten as
\begin{equation}
-\frac{1}{2} \sqrt{PrRa} \tanh{\left(\sqrt{\frac{Ra}{Pr}}\ \frac{z}{4} \right)}  =  \frac{d}{dz} \log{\theta_z} 
\end{equation}
and integrated to
\begin{equation}
\log{\left( \cosh{\left(\sqrt{\frac{Ra}{Pr}}\ \frac{z}{4} \right)} \right)^{-2 Pr}}  =  \log{\theta_z} + \log K 
\end{equation}
 where $\log K$ is a constant, hence removing the $\log$:
 \begin{equation}
  \left( \cosh{\left(\sqrt{\frac{Ra}{Pr}}\ \frac{z}{4} \right)} \right)^{-2 Pr}  =  K \theta_z.
 \end{equation}
 This can be integrated in the interval $[0,z]$
  \begin{equation}
  \theta(z) - \theta(0)= \frac{1}{K} \int_0^z \left( \cosh{\left(\sqrt{\frac{Ra}{Pr}}\ \frac{z'}{4} \right)} \right)^{-2 Pr}  dz'
  \end{equation}
 We now apply the boundary condition $\theta(0) = 0$, while the value of $K$ is obtained by matching the inner and outer solutions for $\theta$, $\lim_{z\to\infty} \theta_{in} =  \lim_{z\to 0} \theta_{out}$:
 $-\frac{1}{2} = \frac{1}{K } \lim_{z\to\infty} \int_0^z \left( \cosh{\left(\sqrt{\frac{Ra}{Pr}}\ \frac{z'}{4} \right)} \right)^{-2 Pr}  dz' $. It follows:
 \begin{equation}
  \theta_{in}(z) =  -\frac{1}{2 }\frac{ \int_0^z \left( \cosh{\left(\sqrt{\frac{Ra}{Pr}}\ \frac{z'}{4} \right)} \right)^{-2 Pr}  dz' }{\int_0^{\infty} \left( \cosh{\left(\sqrt{\frac{Ra}{Pr}}\ \frac{z'}{4} \right)} \right)^{-2 Pr}  dz' }. 
  \end{equation}

The complete perturbative solution is obtained by summing up the inner and the outer solution and by subtracting their overlap, $w_{ma}(z) = w_{in}(z) + w_{outer}(z) - w_{overlap}$, where $w_{overlap} = \lim_{z\to\infty} w_{in} =  \lim_{z\to 0} w_{out} = -\frac{1}{2}$.
 This leads to the final expression for the matched asymptotic solution:
\begin{eqnarray}\label{eq:ma-solution-w}
w_{ma}(z) &=&  z -\frac{1}{2}\tanh\left( \sqrt{\frac{Ra}{Pr}} \frac{z}{4} \right)\\
\label{eq:ma-solution-t}
\theta_{ma}(z) &=& z  -\frac{1}{2 }\frac{ \int_0^z \left( \cosh{\left(\sqrt{\frac{Ra}{Pr}}\ \frac{z'}{4} \right)} \right)^{-2 Pr}  dz' }{\int_0^{\infty} \left( \cosh{\left(\sqrt{\frac{Ra}{Pr}}\ \frac{z'}{4} \right)} \right)^{-2 Pr}  dz' }
\end{eqnarray}
We note that if $Pr =1$ the solutions takes the simpler form
\begin{equation}\label{eq:ma-solution-pr1}
w_{ma}(z) = \theta_{ma}(z)=  z -\frac{1}{2}\tanh\left( \sqrt{Ra} \frac{z}{4} \right).
\end{equation}
The solution (\ref{eq:ma-solution-pr1}) for $Pr=1$ coincides with the shock solution derived by Saffman \cite{Saffman1968} for the randomly forced Burgers equation in the limit of  $t \to +\infty$ and large $Re$ (when $\sqrt{Ra}$ is replaced by $Re$), see also \cite{PhysRevFluids.7.074605} for a recent discussion.

Remark that the matched asymptotic solutions (\ref{eq:ma-solution-w}),(\ref{eq:ma-solution-t}) are valid only in the interval $z \in [0,1/2]$, but they can be applied to the interval $z \in [1/2,1]$ with the transformation $z \to z-1$.
This at the price of accepting a discontinuity in the origin, because e.g. for the velocity $w_{ma}(z=1/2) =\varepsilon \neq 0$. However, such a discontinuity goes as $2\varepsilon=1-\tanh\left( \sqrt{\frac{Ra}{Pr}} \frac{1}{8} \right)$, therefore it vanishes asymptotically with $Ra$.
We also note that the perturbative solution is not an exact solution of the original system. However, it is asymptotically correct. This can be seen by plugging it into the stationary equations and considering the $Ra \to \infty$ limit of the residuals 
\begin{eqnarray*}
&\lim_{Ra \to \infty}& w_{ma} w_{ma,z}  - \tfrac{\sqrt{Pr}}{\sqrt{Ra}}\ w_{ma,zz}  -   \theta_{ma} = 0\\
&\lim_{Ra \to \infty}& w_{ma} \theta_{ma,z} - \tfrac{1}{\sqrt{PrRa}}\ \theta_{ma,zz} -  w_{ma} = 0.
\end{eqnarray*}
 Another observation is in order about the the shape of the solution. We remark here that by choosing that the internal solution occurs at $z=0$, we have implicitly selected the boundary-layer type of the solution. A different, equally admissible choice,  is to place the inner solution close to $z=1/2$, this leads to a shock type solution. One can trivially go from the former type of solution to the latter, by applying the third symmetry transformation (\ref{eq:thirdsym}) discussed in Sec.\ref{sec:sym}.
We observe that these two type of solutions are characterized by the same Reynolds number, as $Re \sim \langle w^2 \rangle^{1/2}$, while they differ for the Nusselt number, because $Nu=1 - \theta_{z}(0)$. The latter observation implies that while boundary layer (BL) type solutions are characterized by an increasing $Nu$ as $Ra\to \infty$, in the shock type solution the Nusselt number vanishes (leading to a perfectly insulating system for $Ra\to \infty$). As we mentioned before, although these two convective states are equally probable, we will limit our considerations to the case of BL type solution, as it offers a better analogy with the dynamics of the realistic RB system which motivates this study.

\subsection{Upper bounds and asymptotic scalings for the Nusselt and Reynolds numbers} 
The matched asymptotic solution allows to promptly compute asymptotic expressions for all global quantities in the system, we focus here on the two main output observables of the BRB system, the Nusselt and the Reynolds number. 

\subsubsection{Nusselt number}
We evaluate $Nu = 1 - \theta_{ma,z}(0)$. 
In the general $Pr$ case, by using (\ref{eq:ma-solution-t}) we get
\begin{equation}\label{eq:Nu-ma}
Nu  = \frac{ 1 }{2 \int_0^{\infty} \left( \cosh{\left(\sqrt{\frac{Ra}{Pr}}\ \frac{z'}{4} \right)} \right)^{-2 Pr}  dz' }
\end{equation}
First of all let us note that the $Ra$  dependence can be factorized by introducing the auxiliary variable $\tilde{z} = \sqrt{Ra}\ z'$ in the integral, so:
\begin{equation}\label{eq:Nu-factorized}
Nu  = \frac{ \sqrt{Ra} }{2 \int_0^{\infty} \left( \cosh{\left(\frac{\tilde{z}}{4\sqrt{Pr}} \right)} \right)^{-2 Pr}  d\tilde{z} }.
\end{equation}
Because the denominator only depends on $Pr$ it is therefore clear that the scaling $Nu \sim \sqrt{Ra}$ is to be expected asymptotically in $Ra$ for any $Pr$.\\
We now focus on  the $Pr$ dependence.
Using  the property $e^{x}/2 \leq  \cosh{x} \leq e^x$ for $x>0$, one can write,
 $$\frac{ 2^{2 Pr +1} }{ \sqrt{Pr} } \geq  
 \int_0^{\infty} \left( \cosh{ \left( \frac{ \tilde{z} }{ 4\sqrt{Pr} } \right) }\right)^{-2 Pr}  d \tilde{z} \geq   
 \frac{2}{\sqrt{Pr}},$$
 and finally by using (\ref{eq:Nu-factorized}):
  $$\frac{\sqrt{Ra Pr}}{4^{Pr+1}}  \leq  Nu \leq 
  \frac{\sqrt{Ra Pr}}{4}.$$
  This bounding relation is relevant in the limit of small $Pr$, because $4^{Pr} \to 1$, leading to the scaling law 
  \begin{equation}\label{eq:nu-pr-small}
  Nu   \simeq  \frac{\sqrt{Ra Pr}}{4}  \quad (Pr\ \textrm{small}).
  \end{equation}
   In the limit of large $Pr$ as $\left( \cosh{\left( \frac{\tilde{z}}{4\sqrt{Pr}} \right)} \right)^{-2 Pr} \to e^{-(\frac{\tilde{z}}{4})^2}$ and using (\ref{eq:Nu-factorized}) one obtains:
  \begin{equation}\label{eq:nu-pr-large}
 Nu  = \frac{\sqrt{Ra}}{4 \sqrt{\pi}}  \quad (Pr \to \infty). 
 \end{equation}
 Which is an expression that does remarkably not depend on the $Pr$ value. It is worth observing that the saturation of the $Nu$ number for large $Pr$ values (\ref{eq:nu-pr-large}) is a feature also observed in the realistic $RB$ system (see e.g. \cite{stevensJFM2013} Figure 5). Finally, for the intermediate case, $Pr=1$, taking advantage of the simpler form of the matched asymptotic solution (\ref{eq:ma-solution-pr1}), one can exactly derive:
\begin{equation}\label{eq:NuRamax}
Nu = \frac{\sqrt{Ra}}{8}  \quad  (Pr=1).
\end{equation}

\subsubsection{Reynolds number}
We now turn the attention to the Reynolds number. Asymptotically in $Ra$ we observe that the $ma$ velocity solution approaches the behaviour $w_{ma}  \simeq w_{out} =  z-1/2$, while the boundary layers (BL) become thinner and thinner. This points to the existence of the upper bound for the velocity variance: $\langle w^2 \rangle \leq \int_{0}^{1} (z-1/2)^2\ dz  = 1/12$, which implies:
 \begin{equation}\label{eq:Remax}
Re = \sqrt{\frac{Ra}{Pr}} \langle w^2 \rangle^{1/2} \leq \frac{1}{\sqrt{12}} \sqrt{\frac{Ra}{ Pr}} 
\end{equation}
A lower bound for $Re$ can be obtained by just considering that the asymptotic outer solution is the one most  contributing to the global velocity variance, so:
\begin{eqnarray}\label{eq:fit}
\langle w^2 \rangle &\gtrsim& \int_{\delta}^{1-\delta} w_{out}^2\ dz \simeq \int_{\delta}^{1-\delta} \left( z - \frac{1}{2} \right)^2 dz\nonumber  = \frac{(1-2\delta)^3}{12}
\end{eqnarray}
where $\delta$ is an estimation for the thickness of the kinetic boundary layer, which we define as the height $z$ where the argument of the hyperbolic function in $w_{in}$  is one, i.e., $\delta = 4/\sqrt{Ra/Pr}$. This implies
\begin{equation}\label{eq:Remax_approx}
Re \gtrsim  \frac{1}{\sqrt{12}} \sqrt{\frac{Ra}{Pr}}  \left(1-8\sqrt{\frac{Pr}{Ra}} \right)^{3/2} 
\end{equation}

We will show that the above predictions for the Nusselt and the Reynolds numbers  approach quite well the result that we obtain from the numerical simulations of the BRB system in the asymptotic large-$Ra$ limit (see  below Sec.~\ref{sec:measureNu}). \\
 
Scaling laws of the form $Nu \sim \sqrt{Ra\ Pr}$ and $Re \sim \sqrt{Ra/ Pr}$ identifies the so called ultimate regime of thermal convection. Physically it can be interpreted as a regime where the microscopic diffusion material properties, i.e. the viscosity and the thermal diffusivities, have a negligible role in the determination of the intensity of the heat transport and of the kinetic energy in the system.
The ultimate regime has been predicted to occur in the RB systems in the asymptotic high-Ra limit  \cite{Kraichnan1962} (see also \cite{grossmann_lohse_2000}). However, its verification in RB experiments and simulations is still debated (see \cite{Doering2020} for a recent concise account). On the opposite, it has been clearly observed in the so called homogeneous-Rayleigh-Benard (HRB) model system, which is a three-dimensional vertically unbounded system, either triperiodic \cite{CalzavariniPF2005} or laterally confined  \cite{Schmidt2012}, which can be realized only in numerical simulations.  It is important to note that the HRB model has no horizontal wall boundaries, as such it misses the corresponding kinetic and thermal boundary layers, i.e. well identified regions where dissipation has the dominant role with respect to inertial transport terms.  \ec{ Experimental realizations of the $Nu \sim \sqrt{Ra}$ regime have been achieved only in bulk dominated convective systems, such as in the vertical channel setup \cite{gibert2006,Arakeri2009} or in systems where the wall thermal heating has been replaced by volumetric radiative heating \cite{Lepot2018,Bouillaut2019}}.
More recently \cite{motoki_2021} have numerically demonstrated the occurrence of the ultimate regime of convection also in a  RB system with permeable walls. This system possesses thermal and kinetic BL but does not enforce the cancellation of the vertical velocity on the top-bottom walls. In the light of this,  the verification of eqs. (\ref{eq:NuRamax}) and (\ref{eq:Remax}) for the BRB system in high-Ra regime would make it the first bounded system, with kinetic and thermal boundary layers, to display the ultimate regime. As we will numerically demonstrate in the next section, the BRB clearly possesses this feature. 
 \begin{figure}[ht]
	\begin{center}
		\includegraphics[width=0.7\textwidth]{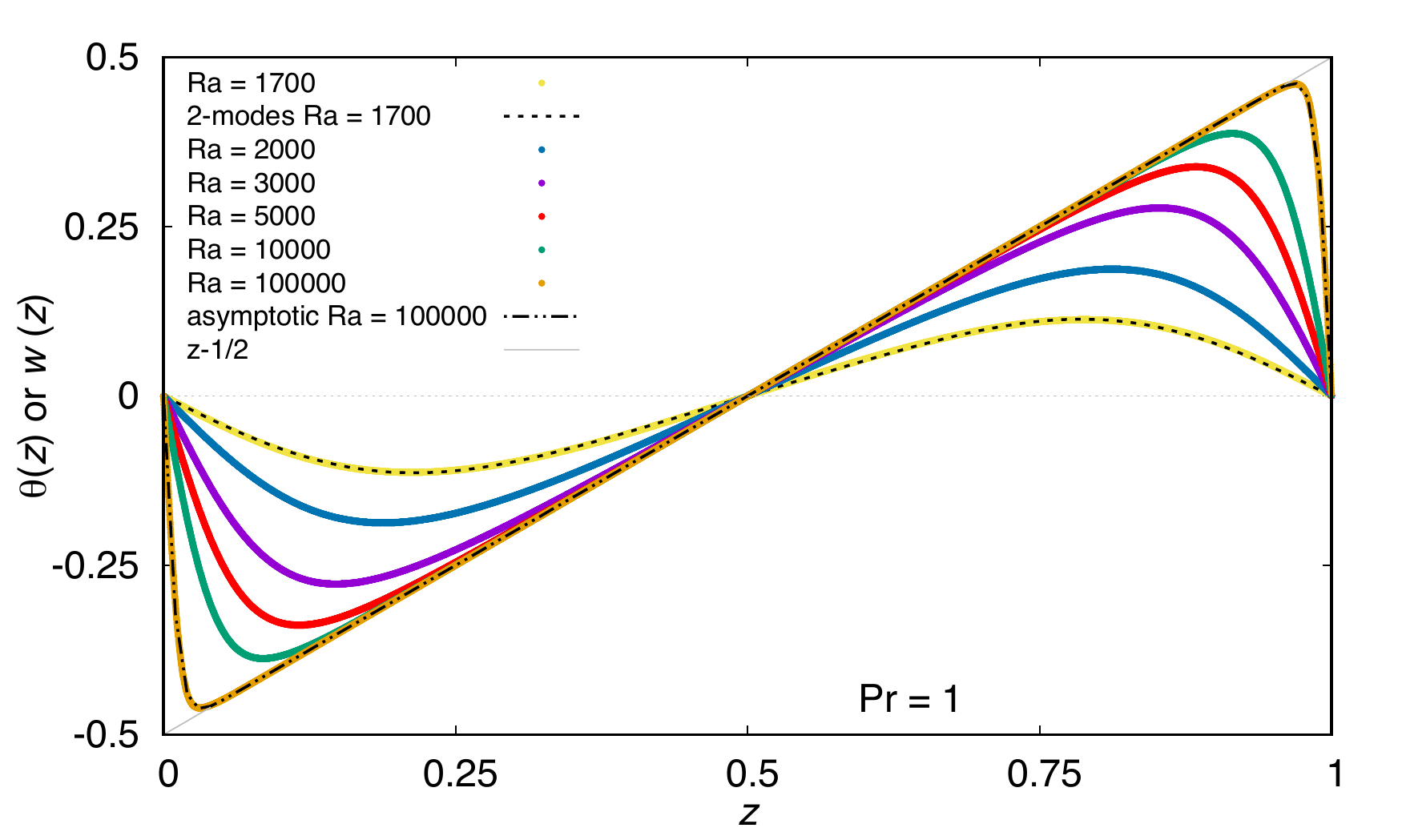}
		\vspace{-0.25cm}
		\caption{Graph of the numerical solution for the temperature fluctuation $\theta(z)$ at $Pr=1$ (identical to the velocity $w(z)$)  for several $Ra$ values from $Ra = 1700 = 1.09 Ra_c$, to $Ra=10^5 = 64.16 Ra_c$ We show the two-modes truncated solution for $Ra=1700$ as well as the matched asymptotic one for $Ra=10^5$. The solution approach the bulk behaviour, $z-1/2$, at increasing the $Ra$ number.}
		\label{fig:sym_pr1}			
	\end{center}
\end{figure}
\begin{figure}[htb]
	\begin{center}
		\includegraphics[width=0.49\textwidth]{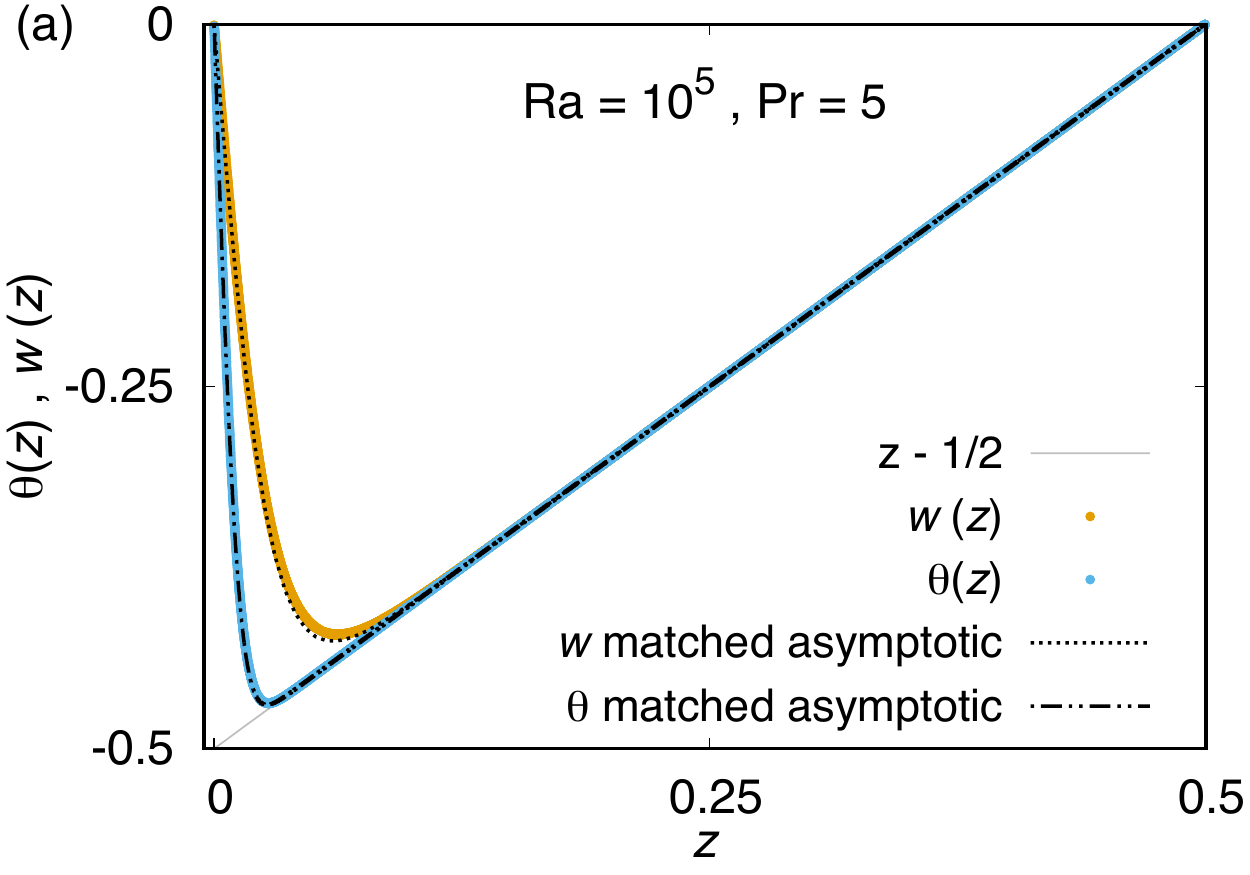}
		\includegraphics[width=0.49\textwidth]{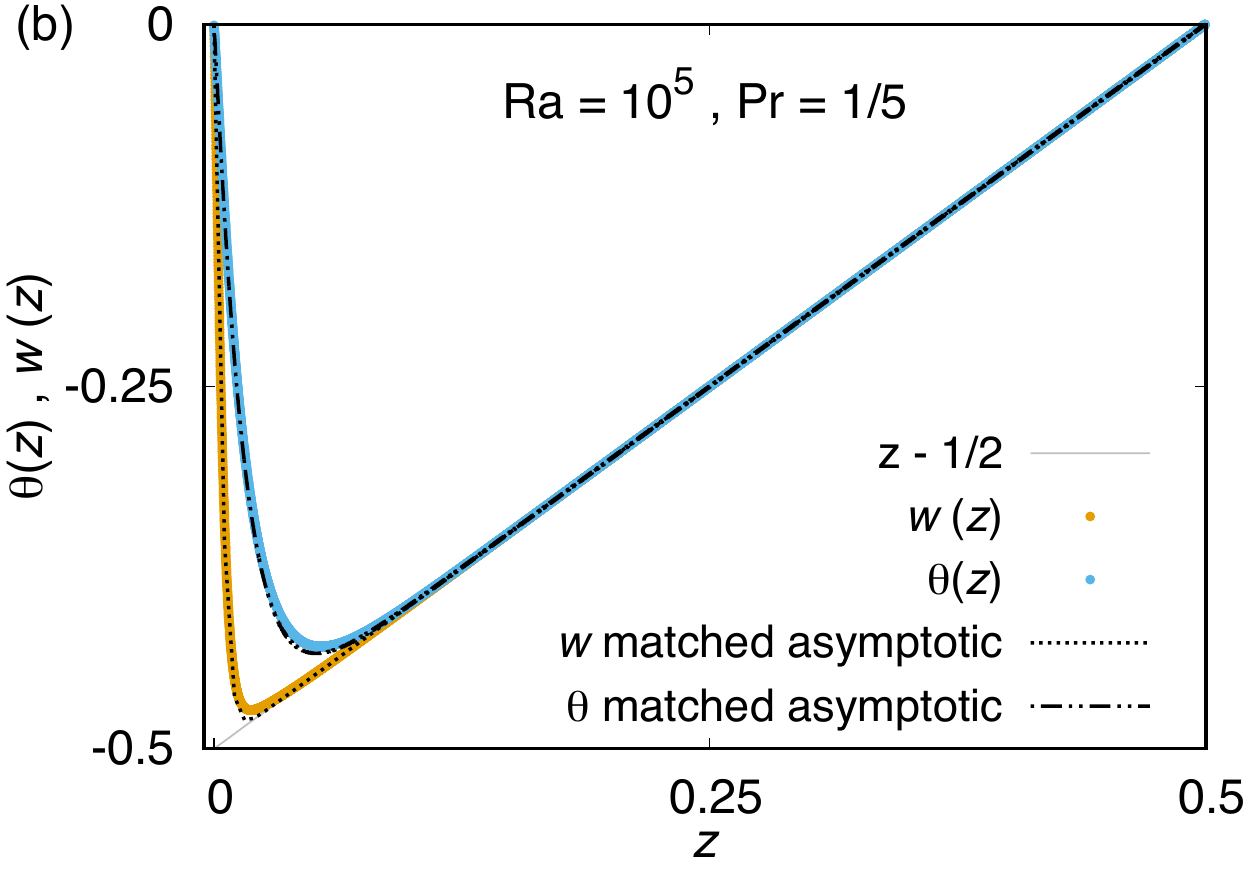}
		\vspace{-0.25cm}
		\caption{Graph of the numerical solution for the temperature fluctuation $\theta(z)$ and velocity $w(z)$ at $Ra=10^5$ and $Pr=5$ (a) and $Pr=1/5$ (b), and comparison with the matched asymptotic solutions for both fields. Note that we show here for better visibility the profile in the bottom half of the domain.}
		\label{fig:sym_pr}			
	\end{center}
\end{figure}	

\section{The BRB dynamics: numerical simulation analysis}\label{sec:numerics}
In order to test the predictions presented in the previous section and to get deeper insight into the BRB dynamics, we performed direct numerical simulations (DNS) of the system of equations. This is conveniently done by means of a Fourier pseudo-spectral method. For this study we use spatial resolutions ranging from $N=2^{13}$ to $2^{16}$ grid points and explore the two-dimensional parameter space $Ra$-$Pr$. The simulations evolve in time from an initial state where $w$ and $\theta$ \ec{are null except for a tiny random uniform spatially-uncorrelated perturbation.  The adopted numerical methods and protocols are described in detail in the Appendix \ref{sec:appendixA}}.  

\subsection{Temperature and velocity profiles}

We numerically find that the system displays a steady solution at any $Ra$, up to $10^{10}$  simulated in this study, and at any $Pr$ in the range $\left[10^{-2},10^2 \right]$. Furthermore, when $Pr=1$ the $w$ and $\theta$ profiles are always coincident.  As illustrated in Fig. \ref{fig:sym_pr1} and  \ref{fig:sym_pr} when the Rayleigh number is slightly beyond the critical threshold $Ra = O(10^3)$, the two-modes solution (\ref{eq:galerkin-coeff1})(\ref{eq:galerkin-coeff2}) perfectly approximate the numerically computed temperature and velocity profiles. On the other hand for $Ra=10^5$ the agreement with the matched asymptotic solution (\ref{eq:ma-solution-pr1}) is already excellent. It is also evident that as $Ra \to +\infty$ the profiles approaches the $z-1/2$ linear shape, with vanishingly thin boundary layers.
Note that the temperature $T$ is the sum of the conductive profile and \ec{the fluctuation} $\theta$, this implies that $T$ is essentially constant in the well mixed bulk of the system and changes sharply only in the boundary layer, in close analogy with the mean vertical temperature profile in the turbulent high-$Ra$ RB system.\\
Figure \ref{fig:sym_pr} reports the corresponding numerical results for non unit Prandtl numbers ($Pr = 5$ and $1/5$) at $Ra=10^5$. One can appreciate that for $Pr>1$ the thermal boundary layer is thinner than the kinetic one and vice-versa for $Pr <1$. Note also that the two cases are distinct, because $Pr \to Pr^{-1}$ is not a symmetry of the system.  Again we observe a good agreement with the matched asymptotic solutions (\ref{eq:ma-solution-w}),(\ref{eq:ma-solution-t}). In the asymptotic high-$Pr$ limit one shall expect that $\theta$ will become almost everywhere equal to $z-1/2$, while in the low-$Pr$ limit the same will happen for $w$.

\subsection{Is the convective stationary regime stable?}
As mentioned above, numerically we find that the convective state of the system displays a stationary solution at any $Ra$ and $Pr$ values. This aspect might seem surprising as it is different from what happens in the RB system, where successive bifurcations occur as $Ra$ is increased, leading first to temporally periodic solutions, then chaotic ones and finally to progressively more turbulent states. In the BRB case, although we can not rule out the existence of sub-critical bifurcations that might lead to the existence of such unsteady states, we can prove that the convective state displayed by the system is linearly stable. This point is addressed in the current section.\\ 

In section \ref{sec:stability} we studied the stability of the conductive state. It was shown that, beyond $Ra_c=k^4=(2\pi)^4$, the system bifurcates to a convective stationary state, hereafter denoted  as $w_s,\theta_s$. Depending on the value of $Ra$ this state can be approximated in three ways: 
\begin{itemize}
\item[(i)] by a two-mode expansion  eq. (\ref{eq:galerkin}) and (\ref{eq:galerkin-coeff1}),(\ref{eq:galerkin-coeff2}) valid near the critical threshold $Ra_c$; 
\item[(ii)] by a matched asymptotic solution  eqs. (\ref{eq:ma-solution-w}),(\ref{eq:ma-solution-t}), valid in the limit of large Rayleigh numbers; 
\item[(iii)] by \ec{an interpolation of} the discretized numerical solution obtained from the DNS\ec{, which is valid in the whole $Ra$ range.}
\end{itemize}
We can now study the stability of these states by applying the linear stability analysis.
We adopt the Galerkin method of weighted residuals \cite{Finlayson1972}. In short, the idea is to choose trial functions that satisfy the boundary conditions exactly and solve the differential equations in an averaged sense, by imposing the condition that the residuals are orthogonal to the trial functions. The result is a set of homogeneous equations whose non-trivial solution leads to an eigenvalue problem. \ec{Here we denote with $\tilde{\sigma}(n)$ the series of the eigenvalues, which are determined in terms of the $Ra$ and $Pr$ parameters.} Similarly to the previous case, we write
\begin{equation}
w=\sum_{n=1}^N \tilde{w}_n \sin(n k z) e^{\tilde{\sigma}(n) t} + w_s , \quad 
\theta=\sum_{n=1}^N \tilde{\theta}_n \sin(n k z) e^{\tilde{\sigma}(n) t} + \theta_s
\end{equation}
with $n\in\mathbb{N}$ and $k=2\pi$. The number of modes $N$ is chosen so that the convergence is assured for the different parameter values. 

\begin{figure}[htb]
	\begin{center}
		\includegraphics[width=0.49\textwidth]{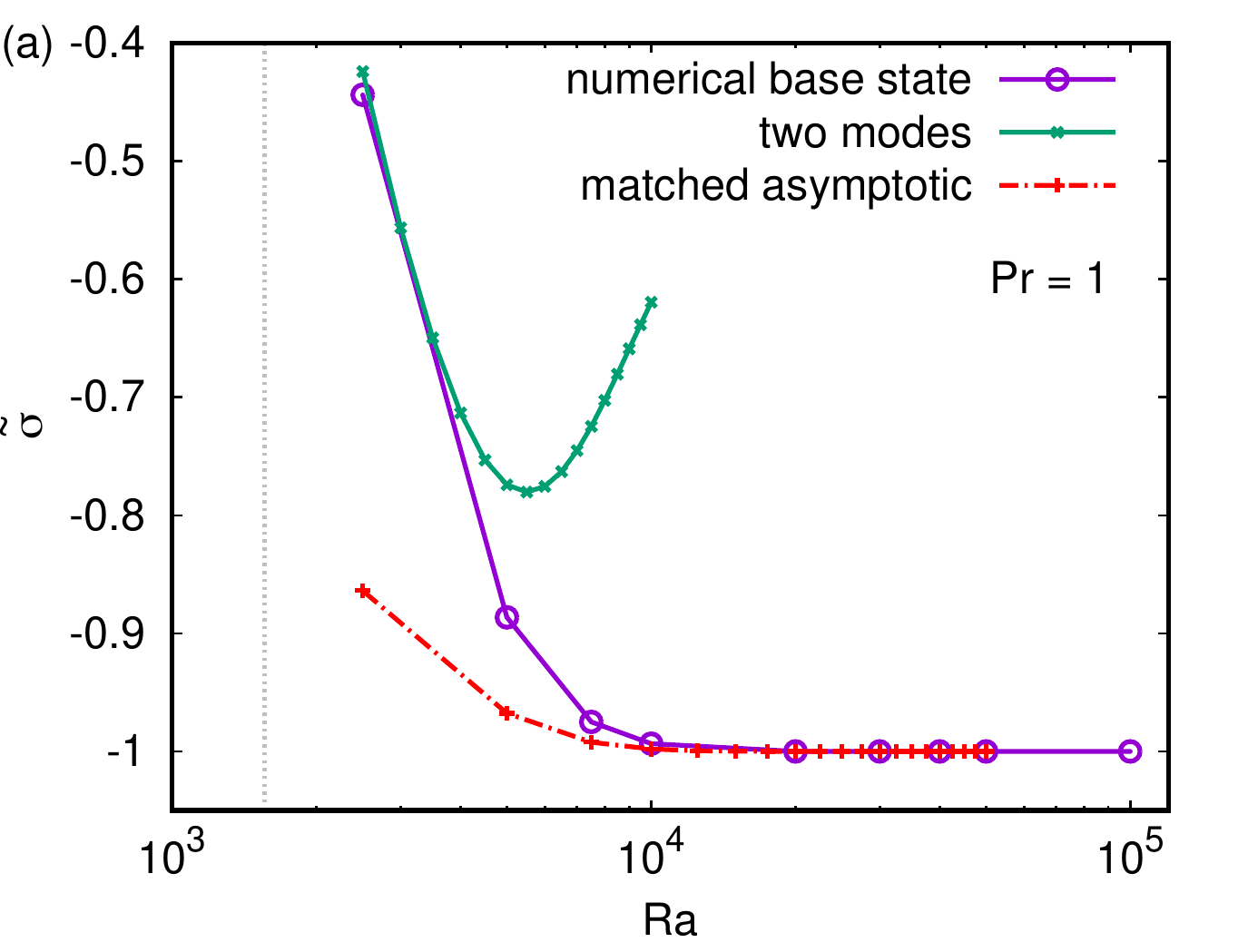}
		\includegraphics[width=0.49\textwidth]{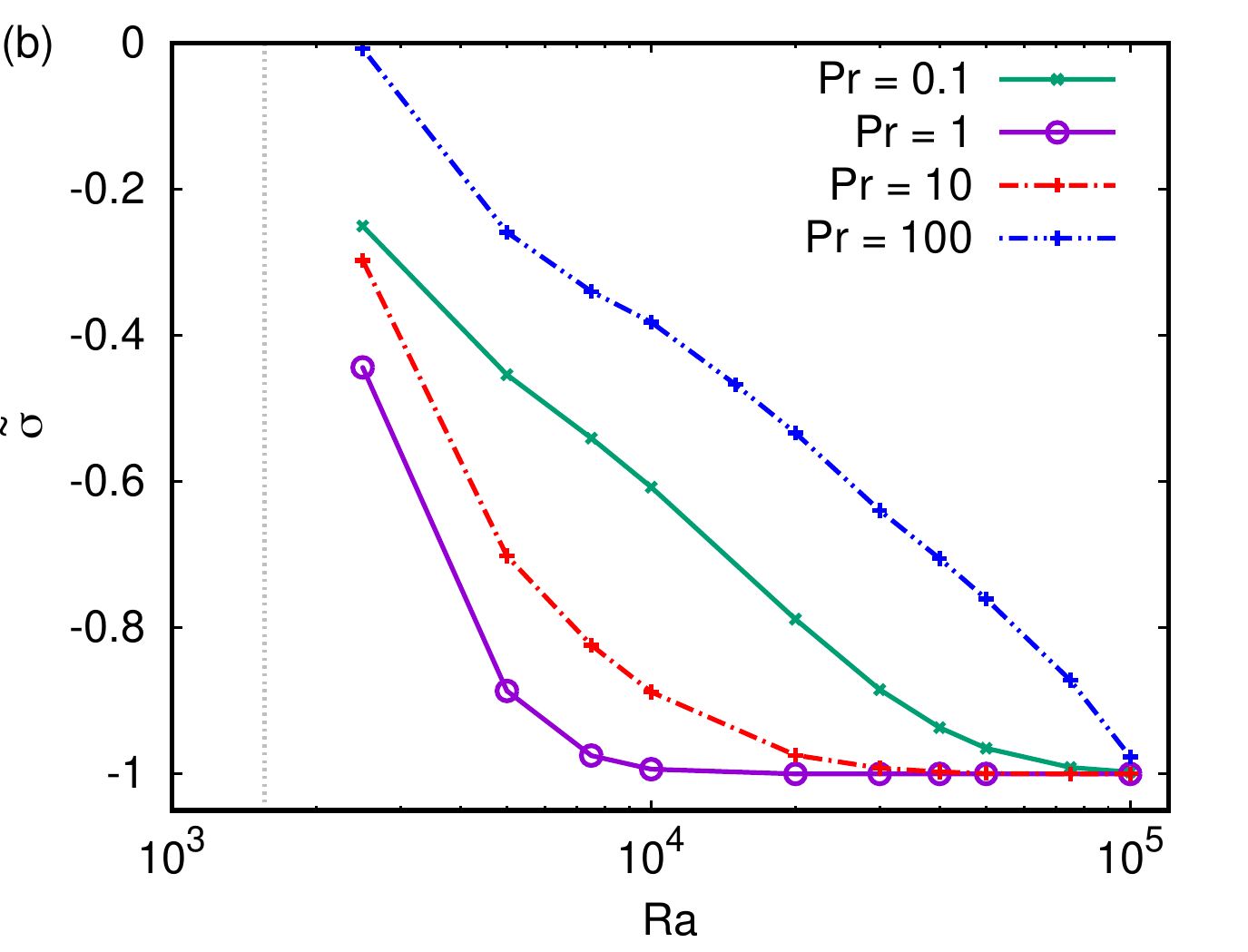}
		\vspace{-0.25cm}
		\caption{Results of linear stability analysis of the convective steady state $w_s,\theta_s$. Perturbation growth rate $\tilde{\sigma}$ as a function of $Ra$: (a) for the case $Pr=1$ computed from three different representation of the base state: the numerical (valid in the full range of $Ra$), the two-mode Galerkin truncation base state (valid at small $Ra$) and the matched asymptotic one (valid for $Ra\to +\infty$); (b) for different Prandtl numbers $Pr \in [0.1,10^2]$ using the numerical base state. The dotted vertical line it is traced at $Ra=Ra_c$.}
		\label{fig:LSA}			
	\end{center}
\end{figure}

Figure \ref{fig:LSA}a shows the evolution of the perturbation growth rate $\tilde{\sigma}=\tilde{\sigma}(1)$ with $Ra(>Ra_c)$, for the three different approximated base solutions and $Pr=1$. As expected, the two-mode solution (case i) agrees well with the numerical convective state (case iii)  for moderate values of the Rayleigh number, while the matched asymptotic does it for large values. The evolution of the larger growth rate $\tilde{\sigma}$ for different values of Prandtl obtained with the numerical base state is depicted on Figure \ref{fig:LSA}b. It can be seen that the growth rate is always negative, indicating that the the first convective stationary state never loses its stability, even for large values of $Ra$, in agreement with the observations from the DNS. 

\subsection{Measure of Nusselt and Reynolds number asymptotic scalings}\label{sec:measureNu}
\begin{figure*}[htb]
	\begin{center}
		\includegraphics[width=0.495\textwidth]{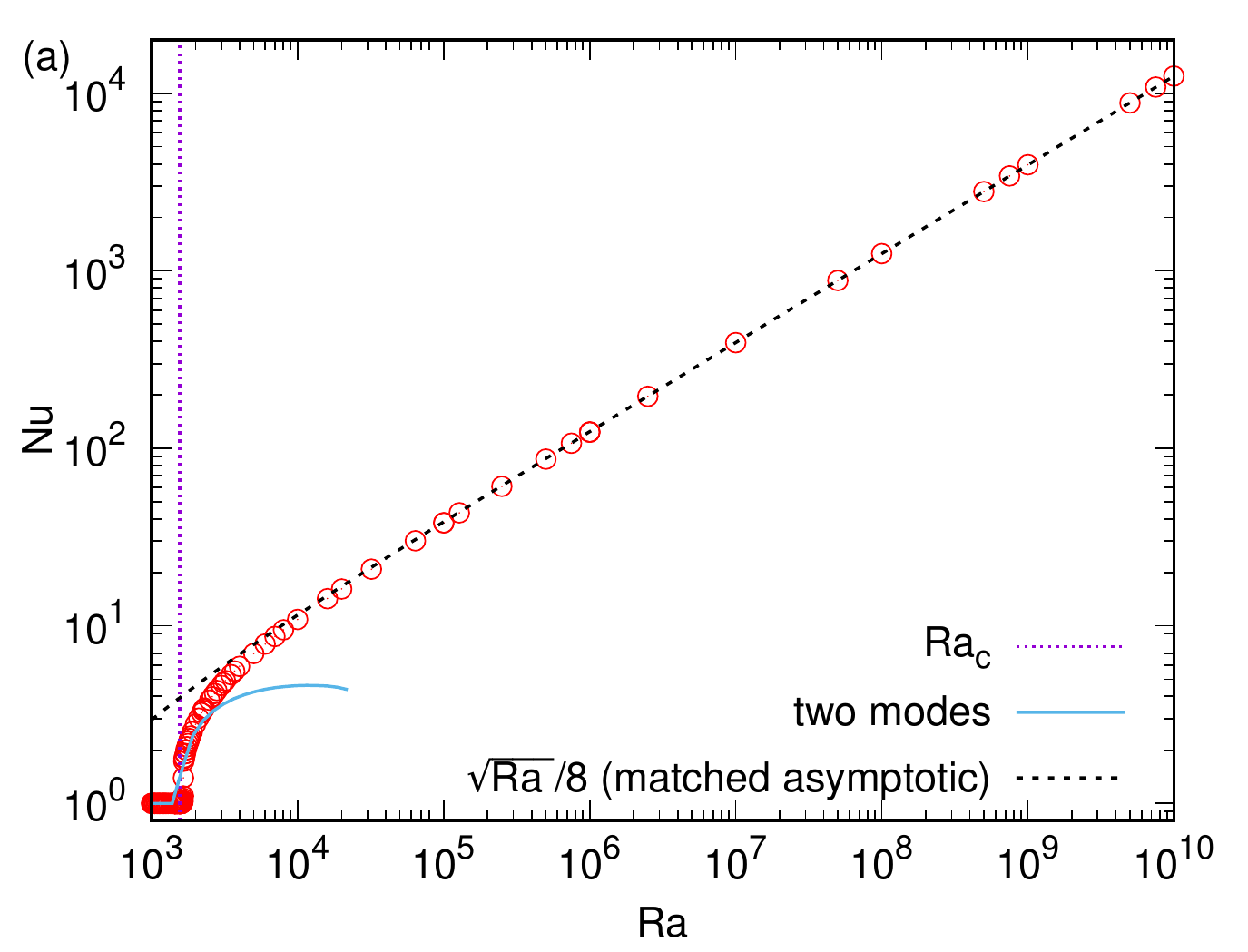}
		\includegraphics[width=0.495\textwidth]{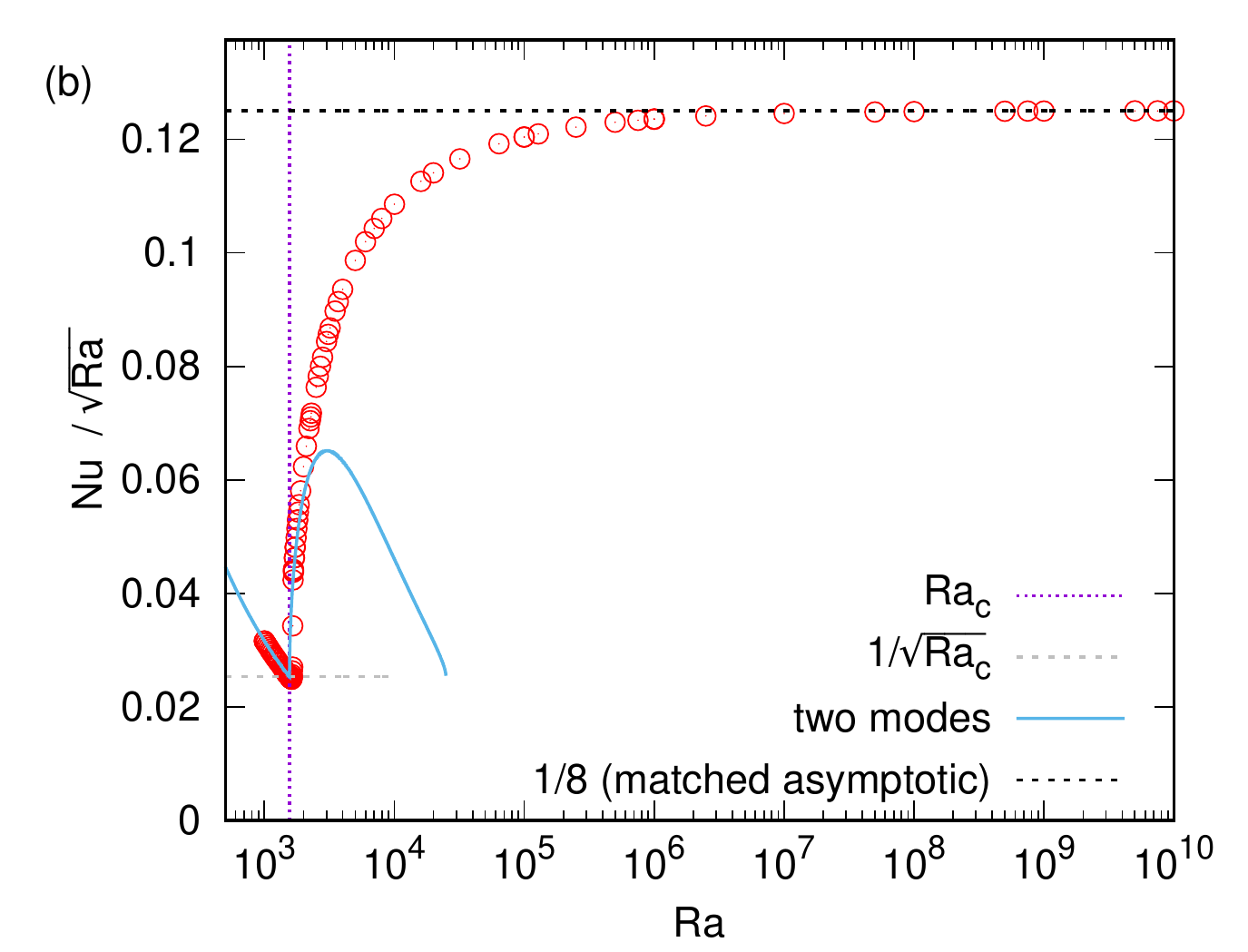}
		\vspace{-0.25cm}
		\caption{Nusselt-Rayleigh scaling at $Pr=1$. (a): $Nu$ vs. $Ra$. We display the numerical simulation results (red circle symbols) and comparisons with different analytical predictions: (vertical dotted line) the critical Rayleigh number for the onset of convection $Ra_c \simeq1558$; (solid line) the Nusselt number calculated from the two-mode steady solution (\ref{eq:galerkin-Nu}); (dashed line) the $Nu$ expression computed via the matched asymptotic solution $Nu =\sqrt{Ra}/8$. (b): The same data points for $Nu$ and theoretical predictions, here compensated with respect to  $Ra^{1/2}$.}
		\label{fig:nu_ra}			
	\end{center}
\end{figure*}

\begin{figure*}[htb]
	\begin{center}
		\includegraphics[width=0.495\textwidth]{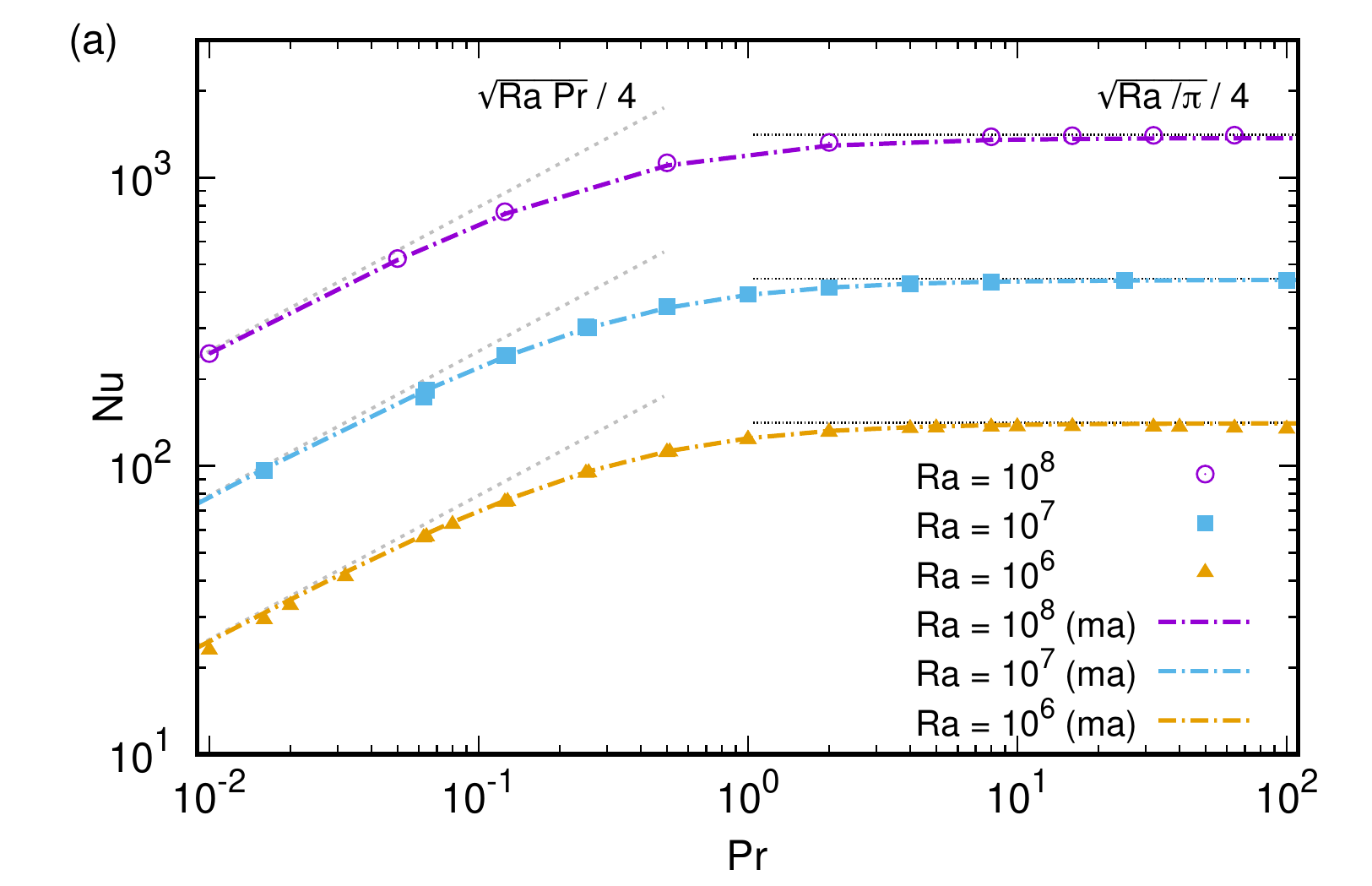}
		\includegraphics[width=0.495\textwidth]{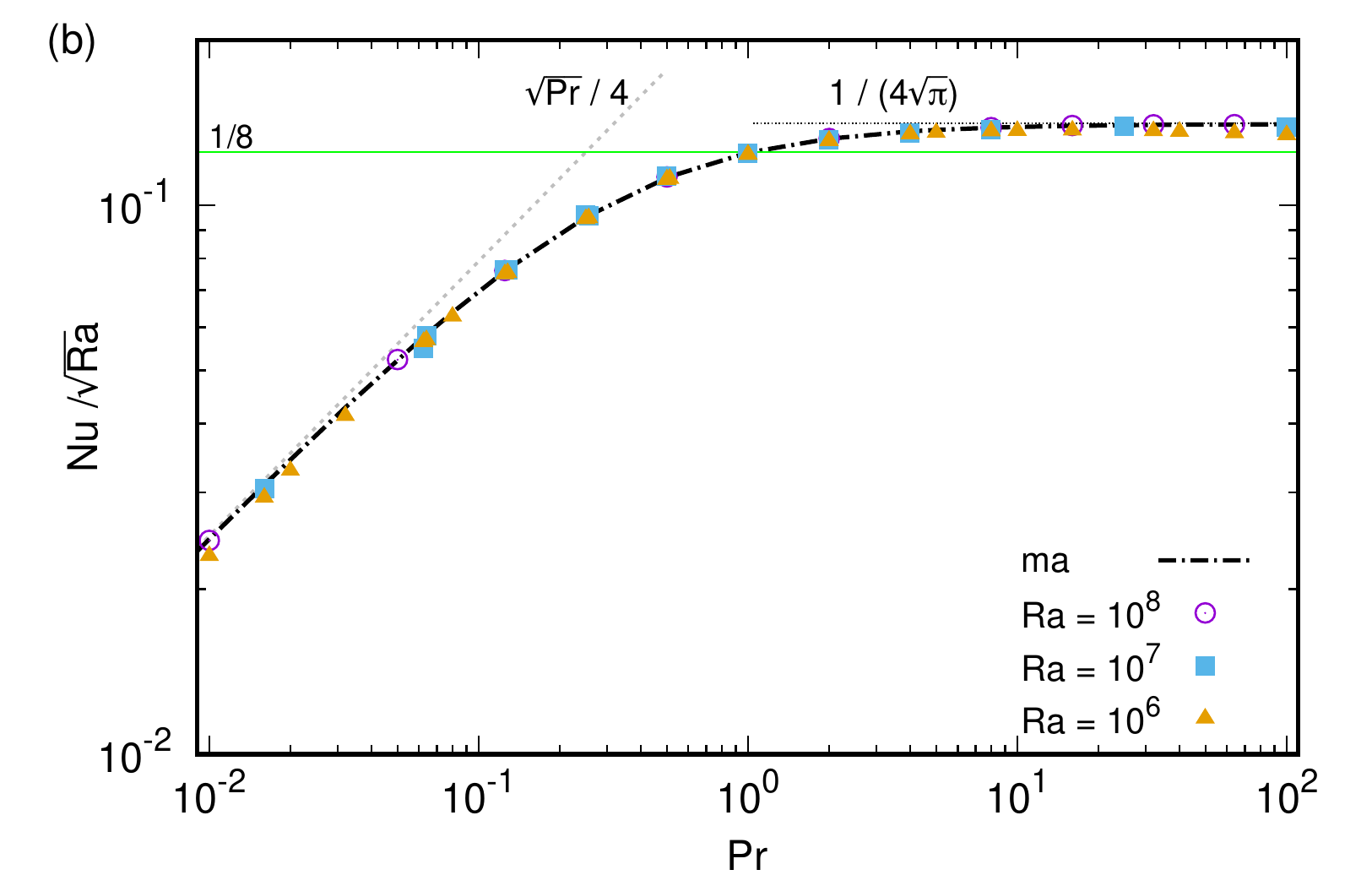}
		\vspace{-0.25cm}
		\caption{Nusselt-Prandlt scaling behaviour at various Rayleigh numbers $Ra=10^6,10^7,10^8$. (a): Nusselt as a function of Prandtl. We display the numerical simulation results (symbols) and the corresponding matched asymptotic predictions (colored dotted-dashed lines). The behaviours $\sqrt{Ra Pr}/4$ for small $Pr$ (grey dashed line) and  $\sqrt{Ra/\pi}/4$ for large $Pr$ (black dotted line). (b): Compensated graph $Nu/\sqrt{Ra}$ vs. $Pr$. The value $1/8$ corresponding to $Pr=1$ is also traced. One can appreciate the collapse of the measurements on a single curve.}
		\label{fig:nu_pr}			
	\end{center}
\end{figure*}

Form the numerically computed $\theta(z)$ and $w(z)$ profiles one can estimate the corresponding Nusselt and Reynolds numbers and analyze their functional dependencies with
$Ra$ and $Pr$.\\
 The behaviour of $Nu$ as a function of $Ra$, for $Pr=1$,  is shown in Fig. \ref{fig:nu_ra}(a). We observe that for $Ra <Ra_c$  $Nu\simeq 1$ (vertical dotted line), hence only conductive heat-transfer occurs, as expected from the theoretical analysis.  Beyond $Ra_c$ convection starts and $Nu$ progressively increases. Close to the onset, for $Ra \lesssim 2 Ra_c \sim 3 \times 10^3$, the two-modes expression (\ref{eq:galerkin-Nu}), (solid line), approaches well the numerical results.  On the other hand the matched asymptotic prediction agrees with the data from $Ra\sim 10^4$ (dashed line) up to the highst explored $Ra$ number ($10^{10}$). Indeed, as it is better appreciated in the compensated plot Fig. \ref{fig:nu_ra}(b) at the largest $Ra$ the normalized data approaches the value $1/8$ (dashed-dotted line) meaning that $Nu \sim \sqrt{Ra}/8$ in excellent agreement with the $ma$ prediction (\ref{eq:NuRamax}).\\ 
We now look at the heat-flux dependence with respect to $Pr$. This is illustrated in Figure \ref{fig:nu_pr}(a), where  $Nu(Pr)$ is traced for various Rayleigh numbers $Ra=10^6,10^7,10^8$. Also in this case we see an excellent agreement with the matched asymptotic solution (\ref{eq:Nu-ma}) and with the small/large-$Pr$ asymptotic behaviours (\ref{eq:nu-pr-small}),(\ref{eq:nu-pr-large}) derived in the previous section. We clearly observe a saturation of the Nusselt number for $Pr \gg 1$.
Furthermore, Figure \ref{fig:nu_pr}(b)  shows how all the data points can be collapsed on a single curve, by means of the rescaling $Nu/\sqrt{Ra}$. This means that, in  agreement with the matched asymptotic solution, the $Ra$ and $Pr$ dependence can be factorized (see eq. (\ref{eq:Nu-factorized})).

\begin{figure*}[htb]
	\begin{center}
		\includegraphics[width=0.495\textwidth]{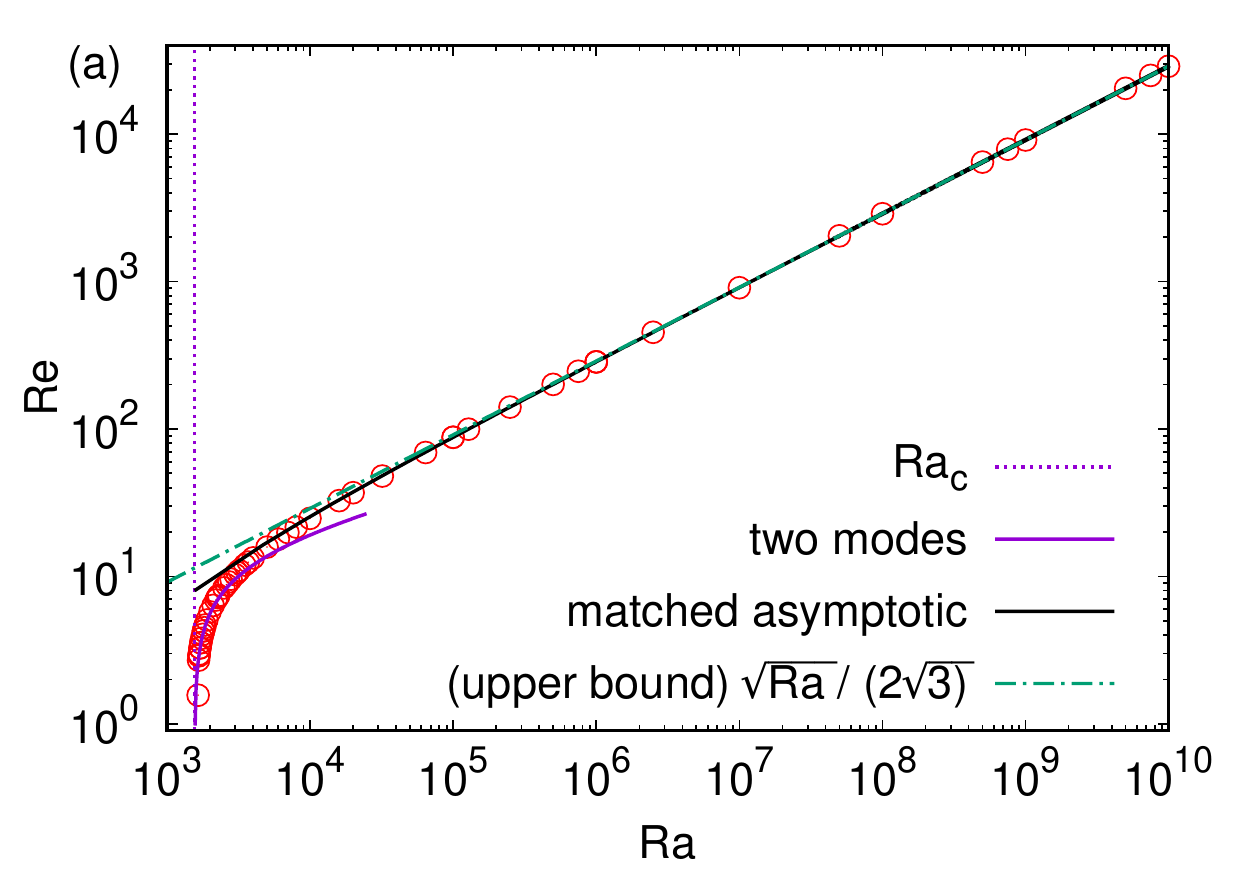}
		\includegraphics[width=0.495\textwidth]{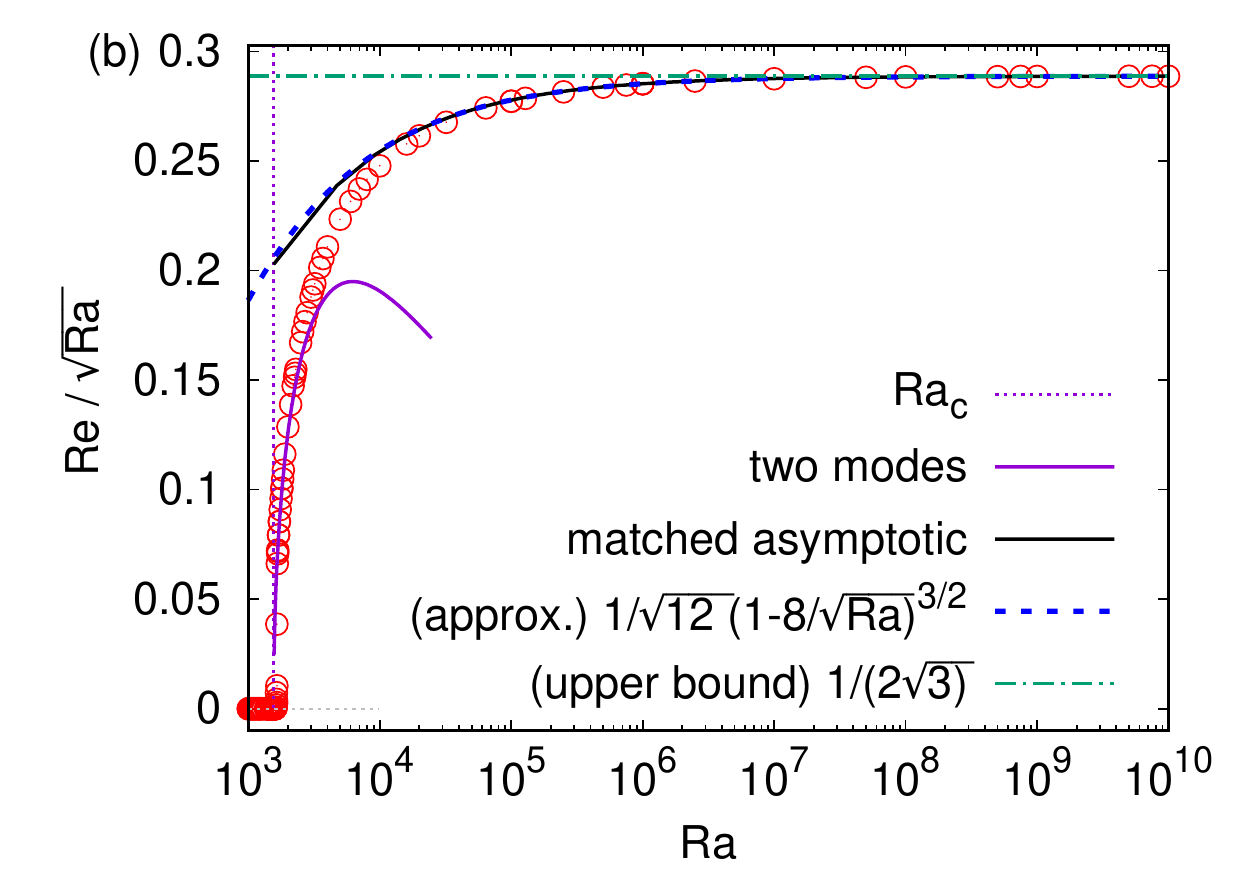}
		\vspace{-0.25cm}
		\caption{Reynolds-Rayleigh scaling at $Pr=1$. (a): $Re$ vs. $Ra$. We display the numerical simulation results (red circle symbols) and comparisons with different analytical predictions: (vertical dotted line) the critical Rayleigh number for the onset of convection $Ra_c \simeq1558$; (violet solid line) the Reynolds number calculated from the two-mode steady solution (\ref{eq:galerkin-re}); (black solid line) the $Re$ expression computed via the matched asymptotic solution; (green dottetd-dashed line) the upper bound value $Re = \sqrt{Ra/12}$ eq (\ref{eq:Remax}). (b): The same data points for $Re$ and theoretical predictions, here compensated with respect to  $Ra^{1/2}$ We include also the approximated expression (\ref{eq:Remax_approx}) (blue dashed line) which has a much simpler form as the $ma$ prediction and fits equally well the measurements at large $Ra$.}
		\label{fig:re_ra}			
	\end{center}
\end{figure*}

\begin{figure*}[htb]
	\begin{center}
	\includegraphics[width=0.495\textwidth]{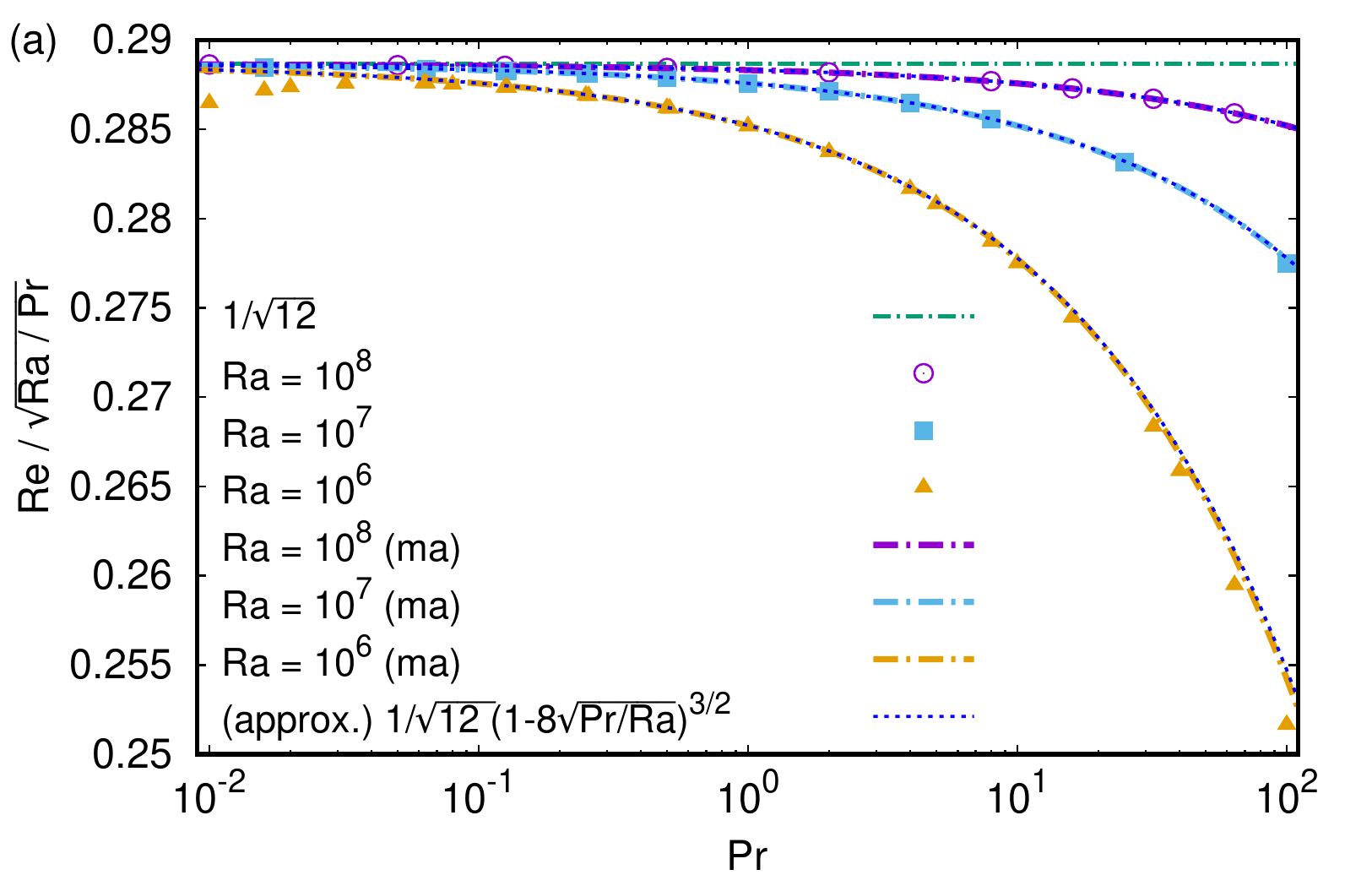}
	\includegraphics[width=0.495\textwidth]{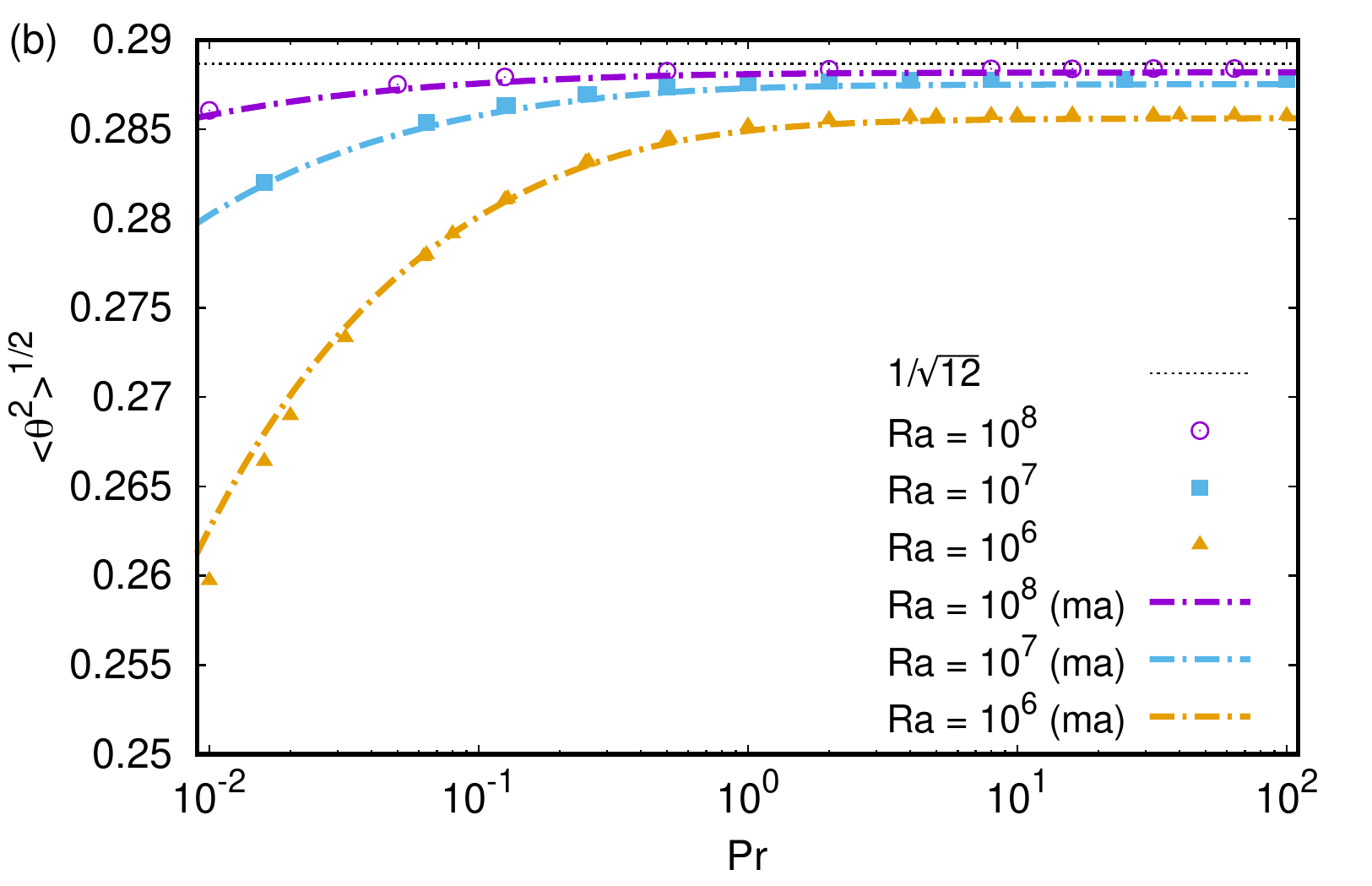}
		\vspace{-0.25cm}
		\caption{Reynolds number scaling versus $Pr$.
		(a) Compensated Reynolds number $Re/\sqrt{Ra/Pr}$, equivalent to the root-mean-square value of the velocity $\sqrt{\langle w^2 \rangle}$ as a function of $Pr  \in [10^{-2}, 10^2]$ for $Ra=10^6,10^7,10^8$; (b) root-mean-square value of the temperature fluctuation $\sqrt{\langle \theta^2 \rangle}$. In both cases we display the numerical simulation results (symbols) and comparisons with the dependencies computed upon integration of the matched asymptotic solutions (\ref{eq:ma-solution-w}),(\ref{eq:ma-solution-t}) and the upper bound $1/\sqrt{12}$ (green dashed-dotted) that can be computed from the bulk solution. In (a) the algebraic approximation  (\ref{eq:Remax_approx}) (dotted) is also reported.}
		\label{fig:re_pr}			
	\end{center}
\end{figure*}

Similar observations can be made for the dependence of the Reynolds number versus $Ra$ for $Pr=1$ (Fig. \ref{fig:re_ra}(a,b)) and $Re(Pr)$ figure Fig. \ref{fig:re_pr} (a)). At high-$Ra$ agreement with the $ma$ predictions is satisfactory in all cases.  The $Re(Ra,Pr)$ functional relation is also well described by the expression (\ref{eq:Remax_approx}), which is simpler in form than the $ma$ expression (see again Fig. \ref{fig:re_ra}(b) and Fig. \ref{fig:re_pr} (a)).
In Figure \ref{fig:re_pr}(a), we observe that the compensated Reynolds number expression $Re/\sqrt{Ra/Pr}$, which is equivalent to the mean root-mean-squared velocity $\sqrt{\langle w^2 \rangle}$, decreases for large $Pr$. This is consistent with the fact that the kinematic boundary layer becomes thicker hence the velocity reduce in intensity. The opposite is true for the fluctuations of the temperature field $\sqrt{\langle \theta^2 \rangle}$,  which is reported in Fig. \ref{fig:re_pr}(b), and for which again the $ma$ solution offers an excellent approximation.

Overall  the DNS confirms the realization of the ultimate regime for the Reynolds and Nusselt numbers in respect to both the Rayleigh and Prandtl dependence. To our knowledge the occurrence of this regime was previously assessed only for the 3D homogeneous-Rayleigh-B\'enard system in \cite{CalzavariniPF2005} \ec{and more recently in a wider $Ra$ and $Pr$ range in \cite{barral_dubrulle_2023}.}
Despite its great degree of abstraction (1D, compressible flow) the BRB system represents a second convective model system where this flow regime takes place. Moreover the saturation of $Nu$ for large-$Pr$ is a feature of the RB model \cite{RevModPhys.81.503} which is here present, while it was instead missing in the HRB system.


\section{Conclusions}
The BRB dynamics shares remarkable similarities with realistic thermal convection in higher spatial dimensions, i.e., the Rayleigh-B\'enard system under Oberbeck-Boussinesq conditions. In this work we have shown that: i) BRB has a supercritical linear instability for the onset of convection which solely depends on the Rayleigh number  and not on Prandlt (same as in RB), occurring at the critical value $Ra_c \approx 1558$ which is of the same order as in the RB system; ii) the convective regime is spatially organized in distinct boundary-layers and bulk regions, although shock like solutions are equally admitted;  iii) the asymptotic high $Ra$ limit displays the ultimate Nusselt and Reynolds numbers scaling regime $Nu = \sqrt{RaPr}/4$ for $Pr\ll 1$, $Nu=\sqrt{Ra}/(4\sqrt{\pi})$ for $Pr\gg1$ and $Re = \sqrt{Ra/Pr}/\sqrt{12}$ thus making BRB the simplest convective system with boundaries exhibiting the ultimate regime of convection. 
A major difference with realistic higher dimensional natural convection is the absence of turbulence. The BRB dynamics is stationary at all $Ra$ numbers above the onset of convection for all $Pr$ values, a feature that results from a nonlinear saturation mechanism.

One may object that the odd symmetry in $w$ and $\theta$ makes the BRB de facto a periodic system, as the fields can be expressed in term of sine series. For this reason in the future it would be interesting to explore the dynamics of this system in a higher (two- or three-) dimensional space.
In this case the convective state might be unstable due to the increased number of degrees of freedom and to the increased system symmetries available in higher dimensions.

Weather the missing physics in the present model, namely the violation of the incompressibility, the absence of the pressure field or of the spatial lateral dimensions, might be related to the realization of the ultimate regime of heat transfer remains a question for further investigations. In this sense it would be interesting, and perhaps useful, to think up of a similar minimalistic model capable to reproduce the classical scaling of thermal convection ($Nu \sim Ra^{1/3}$), in order to see which key mathematical terms and corresponding physical features are needed for the realization of this different convection regime.\\

\paragraph*{Acknowledgments}The authors are grateful to Prof.~M.~N.~Ouarzazi for useful comments.

\appendix

\section{Numerical simulation method}\label{sec:appendixA}
\ec{ Equations (\ref{eq:burgers})-(\ref{eq:theta}) in Fourier space, denoted with $(\, \, \widetilde{} \,\,)$, read:}
\begin{eqnarray}
\tilde{w}_t + \widetilde{w w_z} &=& -k^2 \sqrt{\frac{Pr}{Ra}}\ \widetilde{w} +  \tilde{\theta} \label{eq:fburgers}\\
\tilde{\theta}_t + \widetilde{w \theta_z} &=& -k^2 \frac{1}{\sqrt{Pr Ra}} \widetilde{\theta} +  \tilde{w} \label{eq:ftheta}.
\end{eqnarray}
The dissipative terms can be analytically integrated, while the nonlinear terms can be evaluated via pseudo-spectral algorithm.
In our code \ec{we adopt the standard 2/3 dealiasing procedure for the computation of the nonlinear terms. Furthermore, }we enforce the boundary conditions by imposing that both $w$ and $\theta$ are in real space zero-mean odd-functions (i.e. we use sinus transform instead of Fourier). 
The temporal discretisation with time step $\delta t$, performed by means of a second order Adams-Bashfort algorithm, leads to:
\begin{widetext}
\begin{eqnarray}
\tilde{w}_{n+1}    &=& \left( \tilde{w}_{n} + \frac{\delta t}{2} \left( 3 \left[- \widetilde{w w_z} +   \tilde{\theta}\right]_n - \left[- \widetilde{w w_z} +   \tilde{\theta}\right]_{n-1} e^{-k^2  \sqrt{\frac{Pr}{Ra}} \delta t} \right) \right)e^{-k^2  \sqrt{\frac{Pr}{Ra}} \delta t}\\
\tilde{\theta}_{n+1}   &=& \left( \tilde{\theta}_{n} + \frac{\delta t}{2} \left( 3 \left[ - \widetilde{w \theta_z} +  \tilde{w} \right]_n - \left[ - \widetilde{w \theta_z} +  \tilde{w} \right]_{n-1} e^{-k^2 \frac{1}{\sqrt{Pr Ra}} \delta t}\right) \right) e^{-k^2 \frac{1}{\sqrt{Pr Ra}} \delta t},
\end{eqnarray}
\end{widetext}
where the subscript indexes indicate the discretized value of time. 
The time step width is chosen as $\delta t = 10^{-2} / \sigma$ where $\sigma$ is the growth rate of the most unstable mode, eq. (\ref{eq:sigma}), while the spatial resolution is increased till when \ec{the resulting velocity and temperature profiles become independent}  on the number of discretization points $N$. However, we note that the existence of sharp variations in the solution is at odds with our discretization method based on sine-Fourier transform which is known being affected by from the Gibbs phenomenon. Indeed, at high-$Ra$ some Gibb's like  spurious fluctuations are seen in correspondence of the bulk to boundary layer transition. \ec{This fluctuations do not affect the scaling laws presented in this work.}
The simulations of the BRB system have been also validated against a second code based on finite-difference discretization.\\
\ec{The temperature and velocity fields in the simulations are initialized with a spatially uncorrelated  pseudo-random noise. These perturbations lead in $50\%$ of cases to the BL-type solution and in the remaining cases to the shock solution. Although these two states correspond to different global heat-transfer modes, the resulting velocity and temperature profiles can be transformed one into another by the swap transformation, eq. (\ref{eq:thirdsym}). In the present analysis, focused on the scaling of Nusselt in the BL type state, we take advantage of the swap transformation to maximize the the number of realizations of BL solutions.
However, we note that it is possible to direct the instability towards one of the two possible convective states, e.g. by adding a sinusoidal modulation to the initial white noise. A modulation of the form, $-\sin(2 \pi z)$, leads to BL state while its opposite, $\sin(2 \pi z)$, favours the transition towards the shock type solution. This aspect has an analogous in the RB system where one can control the large-scale circulation direction of the convective cells by intializing the flow with an horizontally asymmetric perturbation of the temperature field.\\}
The codes used in this study are available  at  \url{https://github.com/ecalzavarini/BurgersRB}.


%

\end{document}